\documentclass{article}
\usepackage{psfig}
\usepackage{emulateapj}

\def \sax {BeppoSAX}
\def \degmark{^\circ}
\def \NH {N$_{\mathrm H}$}

\def \hcm {\hbox {\ifmmode $ atom cm$^{-2}\else atom cm$^{-2}$\fi}}
\def \arcmin {\hbox{$^\prime$}}
\def \arcsec {\hbox{$^{\prime\prime}$}}

\def\approxgt{\mathrel{\hbox{\rlap{\lower.55ex \hbox {$\sim$}}
        \kern-.3em \raise.4ex \hbox{$>$}}}}
\def\approxlt{\mathrel{\hbox{\rlap{\lower.55ex \hbox {$\sim$}}
        \kern-.3em \raise.4ex \hbox{$<$}}}}

\slugcomment{ApJ, accepted Nov 5, 2003}

\begin{document}

\title{Non-thermal hard X-ray emission in galaxy clusters observed with the BeppoSAX PDS}

\author{J. Nevalainen$^{1,2,3}$, T. Oosterbroek$^{2}$,  M. Bonamente$^{3}$, S. Colafrancesco$^{4}$}
\affil{Harvard - Smithsonian Center for Astrophysics, Cambridge, USA$^{1}$\\
ESTEC, Noordwijk, Netherlands$^{2}$\\
University of Alabama in Huntsville, Huntsville, USA$^{3}$\\
INAF - Osservatorio Astronomico di Roma, Rome, Italy}

%\email{jukka@head-cfa.harvard.edu}

\begin{abstract}
We study the X-ray emission in a sample of galaxy clusters using the BeppoSAX
PDS instrument in the 20 -- 80 keV energy band. We estimate the non-thermal hard
X-ray cluster emission (HXR) by modeling the thermal contribution from the
cluster gas and the non-thermal contamination from the unobscured AGN in the
clusters. We also evaluate the systematic uncertainties due to the background
fluctuations. Assuming negligible contamination from the obscured AGN, the
resulting non-thermal component is detected at a 2$\sigma$ level in $\sim$50\%
of the non-significantly AGN-contaminated clusters: A2142, A2199, A2256, A3376,
Coma, Ophiuchus and Virgo. The data are consistent with a scenario whereby
relaxed clusters have no  hard X-ray component of non-thermal origin,
whereas merger clusters do, with a 20 -- 80 keV luminosity of $\sim 10^{43-44}$
h$_{50}^{-2}$ erg s$^{-1}$. The co-added spectrum of the above clusters
indicates a power-law spectrum for the HXR with a photon index of
2.8$^{+0.3}_{-0.4}$ in the 12 -- 115 keV band, and we find indication that it
has extended distribution. These indications argue against significant
contamination from obscured AGN, which have harder spectra and centrally
concentrated distribution. These results are supportive of the assumption of the
merger shock acceleration of electrons in clusters, which has been proposed as a
possible origin of the non-thermal hard X-ray emission models. Assuming that the
Cosmic Microwave Background photons experience Inverse Compton scattering
from the merger-accelerated relativistic electrons, and thus produce the
observed HXR, the measured hard X-ray slope corresponds to a differential
momentum spectra of the relativistic electrons with a slope of $\mu$ = 3.8 --
5.0. In presence of cluster magnetic fields this relativistic electron
population produces synchrotron emission with a spectral index of 1.4 -- 2.1,
consistent with radio halo observations of merger clusters. Thus both hard X-ray
and radio observations of merger clusters are consistent with the Inverse
Compton model. The observed slope of HXR is also consistent with that predicted
by the non-thermal bremsstrahlung, which thus cannot be ruled by the fit
to the current data, even though this model requires an extreme, untenable
cluster energetics. Assuming centrally concentrated distribution of HXR, the
data requires a harder slope for the HXR spectrum, which is consistent
with secondary electron models, but this model yields a worse fit to the
PDS data and thus seems to be disfavored over the primary electron
Inverse Compton model.

\end{abstract}


\section{Introduction}
Non-thermal hard X-ray (HXR) emission has recently been observed in several clusters and groups of galaxies with the MECS and PDS
instruments onboard BeppoSAX and with GIS instrument onboard ASCA.
In the case of Coma (Fusco-Femiano et al. 1999), A2256 (Fusco-Femiano et al. 2000), HCG62 (Fukazawa et al. 2001) and A754
(Fusco-Femiano et al. 2003) detection of excess emission above the contribution from the hot ICM is
statistically significant, while marginal evidence is provided for A3667
(Fusco-Femiano et al. 2001) and A2199 (Kaastra et al. 1999).

Most models of the HXR emission involve a population of electrons accelerated in
the cluster medium. A natural source of acceleration in clusters is provided by
merger shocks. In a strong cluster merger event, the electrons are accelerated
to relativistic speeds (e.g. Bell, 1978a,b; Fujita \& Sarazin 2001; Takizawa \&
Naito 2000). The Inverse Compton scattering of Cosmic Microwave Background
photons from the relativistic electrons in clusters then can produce a
non-thermal tail which exceeds the thermal bremsstrahlung emission at energies
above 20 keV (e.g. Sarazin 1999, Blasi \& Colafrancesco 1999). This model has
been proposed as the simplest possible explanation of the HXR properties of
galaxy clusters. If the acceleration is provided by a less energetic merger or
turbulence (e.g. Ensslin et al. 1999), or if there is a high-energy cutoff in
the electron velocity distribution, the resulting electron population is
effectively transrelativistic. In this case, the dominating mechanism in
producing hard X-rays has been proposed to be non-thermal bremsstrahlung (e.g.
Sarazin \& Kempner 2000). However, this solution faces some crucial problems
mainly concerning the large energy injection required by such a mechanism and
the resulting large heating expected (e.g., Petrosian 2001).

In the secondary electron population models the merger shocks and galaxy activity accelerate and inject large quantities of
relativistic protons into the cluster atmosphere.
Most of the relativistic protons can be confined and accumulated in the cluster medium for very long times, comparable with the cluster
age $\sim H_0^{-1}$, and can then produce secondary electrons via proton-proton collisions
(e.g., Colafrancesco and Blasi 1998). The energy losses are balanced by the continuous refilling of the new
electrons produced in situ (i.e. continuously in time and everywhere in space). Thus the resulting HXR spectrum in the secondary models
reflects the electron spectrum right after the acceleration event, while in the primary models the more energetic electrons loose energy rapidly and
the HXR spectrum steepens accordingly with time.

In the present work, we expand the database of cluster hard X-ray emission by studying a sample of clusters observed with the BeppoSAX
PDS. We model the thermal and AGN contributions in the sample in order to obtain estimates
for the non-thermal component. We propagate the modeling uncertainties, as well as the background fluctuation
uncertainties in order to obtain reliable (and somewhat conservative) estimates for the non-thermal component. We
furthermore study the co-added non-thermal hard X-ray spectrum of the sample, in order to investigate the origin of this emission.

We consider uncertainties and significances at 1 $\sigma$ level, and use
H = 50 $\times$ h$_{50}$ km s$^{-1}$ Mpc$^{-1}$, unless stated otherwise.
We define HXR in this paper as 20 -- 80 keV PDS net count rate, after removing the sky background, cluster thermal
component and AGN contamination.

\section{PDS analysis}
The sample consists of all publicly available clusters (as of June 2001) observed with BeppoSAX
whose temperatures are constrained within $\sim$10\% .

\subsection{Data processing}
The observations were processed using SAXDAS 2.2.1. Extreme care was taken during the processing of the
PDS data. We removed  spikes which are caused by charged particles hitting only one of the collimators
using the method described by Guainazzi et al. (1997). The effect of the spike-removal was negligible compared to the statistical
uncertainties.

For A1795, A2163, A2256, A3667 and Coma the observations were divided into several exposures, and A3627 has several available
pointings. The datasets were processed separately, and the resulting spectra were co-added.

In order to improve the S/N we restricted the PDS analysis to 20--80 keV band and binned the data to contain a single bin covering
this band.

\subsection{Background subtraction}
\label{bkg}

The total background in 20--80 keV band is $\sim$10 c/s (Frontera et al. 1997b). Because the clusters in our sample
have lower count rates than the background, we addressed carefully the uncertainties involved in the background subtraction.
In the standard observing mode, the PDS system of two identical collimators is rocked back and forth after each 96s, keeping
one collimator at all times pointed at the X-ray target, to allow the simultaneous monitoring of source and background
(Frontera et al. 1997a). In the standard data processing, both offset-positions (3.5$^{\circ}$ away from the source) are used for the
background subtraction. The allowed upper limit for the background modulation with the rocking collimator offset angle is only 2\%
(Frontera et al. 1997a).
Possible variations in the cosmic ray induced internal background due the changing environment are addressed by the
standard method of subtracting the simultaneous background spectrum obtained from regions close to the source.
The hard X-ray sky background is composed mostly of discrete sources (absorbed AGNs) and is not uniform
(e.g. Comastri et al.\ 1995). Fluctuations in the number and flux of the sources could give rise to a significant uncertainty in the
background subtraction. Related to this is the effect of the presence of weak sources in only one of the
offset-positions which is used for the background subtraction (the same effect as described above, except that
the strength of
the source is such that it is discernible in one pointing).
In order to estimate the effect
of random faint sources (e.g. weak AGNs) in the FOV in the offset positions, we determined the count rates for the cluster using
only one of the two (either the negative or positive direction). A clear difference between the spectra could indicate the presence of
source(s) in the offset position. Note that when using only one offset position for the background the exposure time of the background is
effectively halved, which results in a larger uncertainty in the source fluxes.

Initially, in order to test the robustness of the standard method in the case of no background fluctuations, we used a small sample of
12 pointings (Polaris, Lockmann hole, secondary pointings) in which no sources (other than ``background'') are thought to be present.
From this we concluded that the mean 20-80 keV flux is very close to zero (-$(0.4\pm 3.7) 10^{-2}$ c s$^{-1}$) when using the standard
background subtraction, indicating that the systematic uncertainty of the standard background subtraction is negligible compared to
statistical errors (Table \ref{results.tab}). However, when using either the positive or negative offset directions only we noticed a
systematic
difference of $\sim$0.06 counts s$^{-1}$ between these two with the positive pointings giving higher background subtracted
source count rates.

\begin{figure}
\plotone{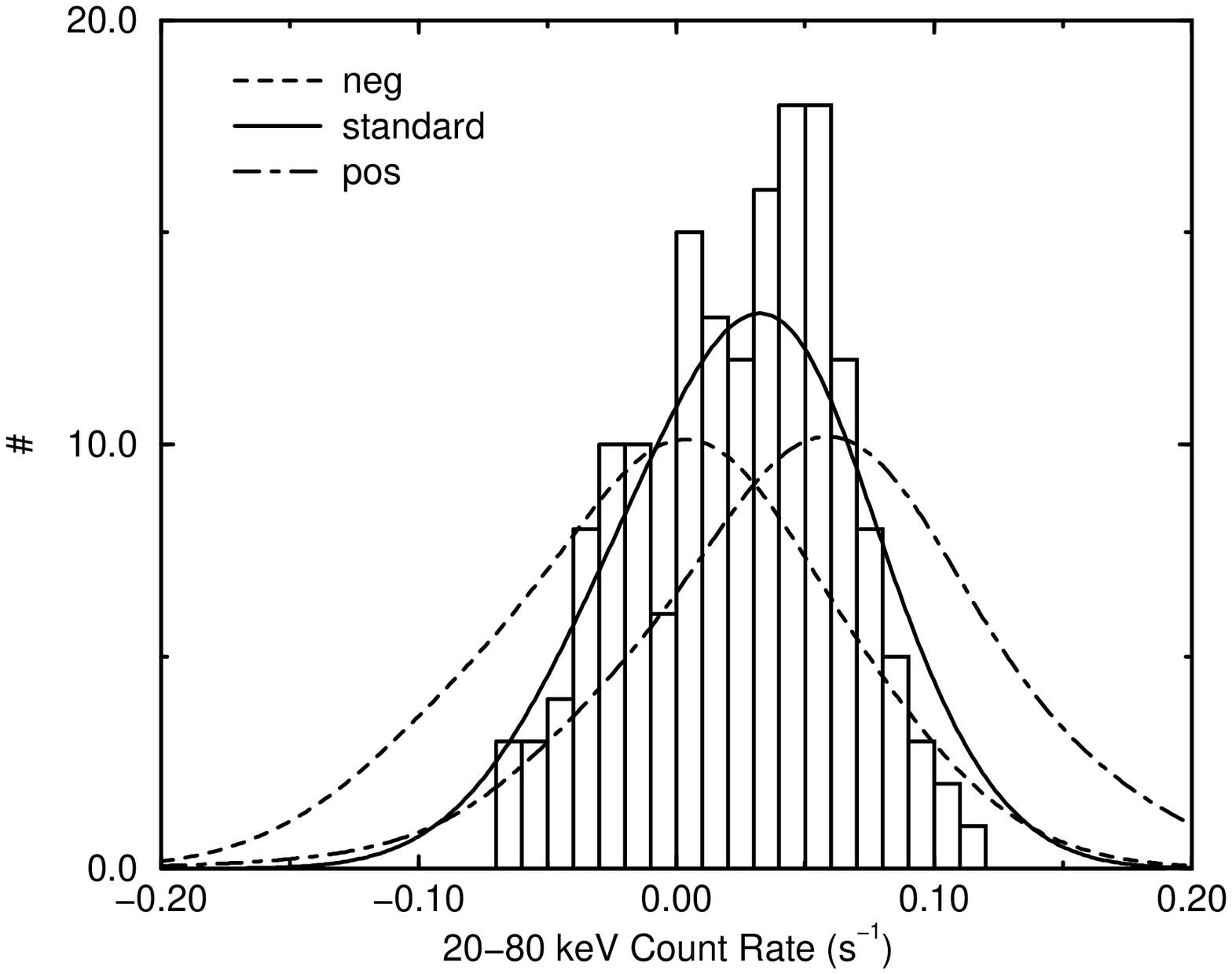}
\caption[]{The PDS count rate distribution of the selected 164 fields in 20--80 keV band.
The histogram shows the data when using the standard method of subtracting the background of both off-sets.
The solid curve is a sum of gaussians which represent the data points with
the value of the data point as the centroid and the uncertainty as $\sigma$.
The dash-dot and dashed lines show the corresponding sum of gaussians, when using only positive or negative off-sets for background
subtraction.}
\label{f1.fig}
\end{figure}

In order to study the above difference due to different offset pointings in more detail we selected
a sample of 164 pointings for which
the (non-spike filtered) exposure times were larger than 20 ks (in order to have similar exposures to the cluster sample) and for which
the count rate obtained with the standard processing (including spike filtering) is between -0.1 and 0.1 counts s$^{-1}$.
These criteria were chosen in order to select either blank fields, or faint sources.
The mean exposure
time ranges between 19.1 ks and 78.3 ks with a mean of 32.2 ks. The spatial distribution of the analyzed pointings is rather uniform.
We find (see Fig.\ \ref{f1.fig}) that the mean count rate as obtained through the standard analysis is $0.027\pm0.051$
counts s$^{-1}$, which suggests a non-significant detection in the {\it whole} sample. The difference between the count rates
obtained with either the negative or positive offset direction amounts to 0.058 counts s$^{-1}$ and the distributions are
symmetric about the mean obtained with the standard analysis. We therefore use a correction of 0.029 counts s$^{-1}$ (with the proper
sign) whenever we use the count rates obtained using only one offset position.
A possible reason for this effect could either be the effect of radiation entering the collimators from the side, screening of the
instruments by the satellite, or the fact that the detector is looking at more/less radioactive parts of the satellite.

Furthermore we quantified the effect of the background fluctuations. To do this, we used the above sample of 164 pointings to compute
the differences in background subtracted count rates between positive and negative background pointing directions. We divided them
into 4 exposure time bins (23, 35, 44, 60 ks) and determined the widths ($\sigma$) of the obtained distributions. The widths
decrease with increasing exposure time, as expected since the widths of the distributions can be described by
$\sigma^{2} =  \sigma_{\rm stat}^{2} + \sigma_{\rm fluc}^{2}$, where $\sigma_{\rm stat}$ is the statistical uncertainty (dependent on
the total number of counts in the background or the exposure time), and $\sigma_{\rm fluc}^{2}$ is due to the real background
variations. We then assumed that $\sigma_{\rm stat}$ is proportional to $\sqrt{t}$ (where $t$ is the exposure time) and fitted this
function, obtaining $\sigma_{\rm fluc}$ = 0.027 cts s$^{-1}$. In the standard background subtraction the uncertainty introduced by the
background fluctuations should be lower by a factor $\sqrt{2}$, i.e., 0.019 cts s$^{-1}$, since 2 background fields are used. In the
following analysis we use this value as a systematic error in the background subtracted source count rates when using the standard method and the
above value 0.027 cts s$^{-1}$ when only one background pointing direction (positive or negative) is used.

\begin{figure}
\plotone{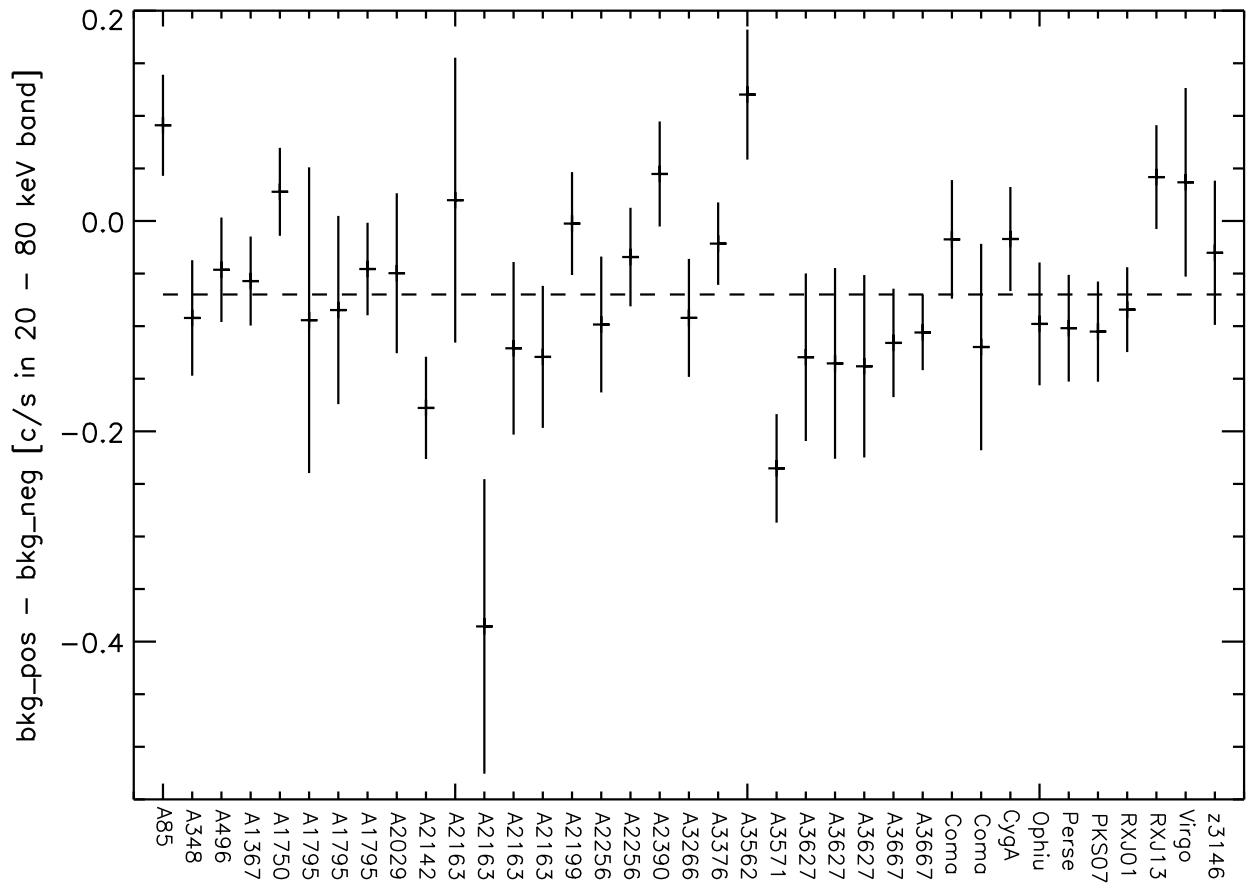}
\caption[]{The  20--80 keV count rate difference between positive and negative offset background pointings for each cluster exposure. The
dashed line shows the average difference.}
\label{f2.fig}
\end{figure}

In Fig. \ref{f2.fig} we plot the difference between the count rates of the two different background pointings for all cluster
exposures. The data give an average difference of 0.068 c s$^{-1}$, very close to the value found in the blank fields above.
Eight exposures deviate from the mean at 90\% confidence level, while random fluctuations would predict only 4, thus our sample is
likely contaminated by point sources in background regions. Thus, we reject the $>$90\% deviant pointings for clusters
A85, A1750, A2142, A2390, A3562, A3571, and RXJ1347.5-1145, and correct the resulting data values by the systematic shift
$\pm$0.029 c s$^{-1}$ found above using the blank fields (for A2163 only a fraction of the total exposure is affected and thus we
make no correction for it). This removes most of the negative count rate detections, with a notable
exception of A3571. The signal of A3571 in 20--80 keV band remains negative regardless of whether we use one offset pointing or
two, indicating that possibly both background offsets are contaminated by AGNs. We chose to use the positive offset for background
subtraction for A3571, because this gives a net count rate closest to zero and thus minimizes the oversubtraction due to AGNs.
For the rest we use the standard method of using both positive and negative background pointings. We add the systematic errors due
to the background fluctuation found above in quadrature to the statistical uncertainties of each cluster. The obtained count rates
are listed in Table \ref{results.tab}.
Using the total 20 -- 80 keV band emission, we achieve 3$\sigma$ detections in the direction of the following 10 clusters:
A2142, A2256, A3376, A3627, A3667, Coma, Cygnus A, Ophiuchus, Perseus and Virgo.

\subsection{Vignetting}
The vignetting of PDS is assumed not to vary with photon energy (Frontera et al. 1997). It is modeled with a linear function of off-axis
angle, reaching zero at 1.3$^{\circ}$. When predicting the thermal contribution at 20 -- 80 keV energies, we use the actual PDS data
(see Section \ref{ther}) at 12 - 20 keV to normalize the model and thus the vignetting is taken care of without further work. When
predicting the AGN contribution, we correct it by multiplying by the vignetting factor at the off-axis of the AGN.
When deriving the non-thermal hard X-ray luminosity for a given cluster, we assume that the HXR is distributed like the intracluster gas
and thus we use the vignetting function and the $\beta$ model to obtain the vignetting correction to the luminosity
obtained with the on-axis response. The effect of the vignetting correction to the luminosity is small, at the level of at most 20\% for
the closest clusters, and thus the assumption on the spatial distribution of HXR is not important.

\section{Thermal models}
\label{ther}
We model the thermal component with XSPEC model wabs $\times$ mekal.
The BeppoSAX MECS study of most of these clusters reports cluster average temperatures (deGrandi et al. 2001). Since in that work the
central regions, affected by the presence of cooler gas (the "cooling flow" scenario)
are excised from the estimates while they are included in PDS data, we preferred to use the published ASCA single
temperature fits (Markevitch et al. 1998) for the "cooling flow" clusters, and BeppoSAX values for the "non-cooling flow" clusters. For the
distant (z$>$0.1) clusters only BeppoSAX results are available and thus we use these regardless of the presence of any cooler gas.

MECS and ASCA do not cover the full FOV of PDS. Thus, due to the radially decreasing temperature profiles consistently observed with
BeppoSAX (deGrandi \& Molendi 2002) and ASCA (Markevitch et al. 1998), the temperatures obtained by these instruments are high compared
to the global ones. For this reason we compare our adopted temperatures with results obtained for a subsample, with Ginga, because it
covers the full FOV of PDS. Ginga temperatures for
A496, A1795, A2142 and A2199 (White et al. 1994) are systematically, and in most cases significantly, below the ASCA values,
consistently with the radially decreasing temperature profiles. However, the thermal model predictions in PDS 20 -- 80 keV band using
either values are consistent within the thermal model normalization uncertainties and thus in general the radial temperature decrease in
clusters  does not significantly affect the conclusions of this work.

For the nearby clusters Coma, Ophiuchus, Perseus and Virgo, the fraction of the cluster covered by MECS and ASCA is small
and thus the effect of decreasing temperature profiles may be significant for these. For example, the $\sim$ 1${\degmark}$ FOV Ginga
temperatures of Coma (8.21$\pm0.16$ keV, Hughes et al. 1993) and Perseus (6.33$^{+0.21}_{-0.18}$, Allen et al, 1992) are smaller than the corresponding
MECS 0-20$'$ values (9.20$\pm$0.13 keV deGrandi et al. 2002 and 6.68$\pm$0.08, respectively). The model prediction in 20 -- 80 keV
band of PDS with Ginga parameters for Coma is significantly smaller than with BeppoSAX values. Thus, for the nearby Coma
and Perseus clusters we adopt the Ginga results. For the Virgo cluster Ginga data are not available and we use the MECS to estimate the
temperature (see Section \ref{agn}). For Ophiuchus there are  $\sim$ 1${\degmark}$ FOV Tenma results (11.6 keV; Okumura et al. 1998) but
these are at odds with the 0$'$-8$'$ MECS value of 10.9$\pm$0.3 keV. Using the Tenma value, and
normalizing the model to PDS 12 -- 20 keV (see below), the model prediction is significantly above the observed emission, indicative of
overestimation of the temperature. For a hot and nearby cluster like Ophiuchus, the PDS data are
of sufficiently good quality for the purpose of spectral fitting. In
the Ophiuchus field there are no contaminating AGN (see Section \ref{agn}) and thus we can assume that the low energy band of PDS
(12--35 keV) is dominated by the thermal emission of the whole cluster. Fitting this band with mekal, keeping the metal abundance fixed to
0.3 Solar, we obtained a temperature of 9.1$\pm$0.6 keV. MECS and PDS values are consistent with the decreasing temperature profile (and
inconsistent with Tenma values), and thus we adopt our PDS results for the thermal model of Ophiuchus.

In order to normalize the above models to the larger (radius of 1.3$^{\circ}$) FOV of PDS, we fitted the thermal models to the
12 -- 20 keV band PDS data with the normalization as the only free parameter. To check the robustness of the fit, we predicted the
normalization using  $\beta$ models to compute the increment of the model normalization between the region where the cluster model has
been normalized, usually by ROSAT PSPC (Ebeling et at. 1996), and the PDS FOV. The fitted and predicted normalizations differ by more
than 50\% for the faint clusters (A348, A1750, A2390, PKS0745-191, RXJ0152.7-135.7 , RXJ1347.5-1145 and Zw3146),
probably due to large statistical uncertainties in the 12--20 keV data. Thus we reject these clusters from further analysis. For the rest
these two methods give values that differ by less than 40\%, which can be explained by the uncertainties involved in the radial
extrapolation and the cross-calibration uncertainty between PDS and PSPC. We prefer to use the fitted values, because they should
be devoid of these uncertainties.

In Virgo, Perseus and Cygnus A the 2--10 keV band data are strongly contaminated by AGNs. Thus, when normalizing
the thermal model, the AGN contribution must be taken into account. We will describe this in detail
in Section 4.

As a further check of the bright AGNs in background fields, we compared the obtained thermal model normalizations in 12 -- 15
and the 15 -- 20 keV energy bands. The appearance of an AGN in only one background field would result in different normalization
using either positive or negative pointing for background subtraction. Also, if the spectrum of an AGN is
not identical to that of the cluster in the 10 -- 20 keV band, its appearance would result in different
normalization using either the 12--15 or 15--20 keV band.
We found that in the 15--20 keV band in all clusters, except A3571, both offsets give consistent values for the
normalization of the thermal model. In the 12--15 keV band clusters A3571 and A2142 give inconsistent values
between the two offsets. Therefore, for A2142 we use only the 15--20 keV band for the normalization, and for A3571
we use the predicted normalization.

The obtained model predictions are listed in Table 1. The reported uncertainties of the thermal models include
only the statistical uncertainty due to the PDS data in 12 -- 20 keV band. In most cases this is
negligible compared to the PDS 20-80 keV count rate uncertainties. We previously discovered that for a subsample of
clusters possible hot ICM temperature variations result in negligible variation of the thermal model
prediction in the 20 -- 80 keV band.
Assuming that this holds for the whole sample, propagating the model normalization uncertainty only is
adequate to estimate the uncertainties in the thermal model.

\section{AGN}
\label{agn}
In the large field of view of PDS the 20--80 keV band emission may be contaminated by AGN and QSO
randomly projected in the line of sight, or AGN belonging to the cluster under study.
In optical surveys it was found that most of the nearby AGN ($\sim$ 80\%) are optically faint Seyfert 2 galaxies
(e.g. Maiolino \& Rieke 1995). In the unified AGN scheme the optically bright and
identifiable AGN or Seyfert 1 (Sy1) are the ones observed face-on with no obscuration by the torus. Most of the lines of sight to the AGN
nucleus are intersected by the absorbing torus and thus most of the AGN are obscured (NH = 10$^{22-25}$ atoms cm$^{-2}$, Risaliti et al.
1999) and optically faint Seyfert 2 (Sy2).
Recent deep X-ray observations of blank fields (e.g. Hasinger et al. 2001) have consistently discovered a population of absorbed point
sources which outnumbers the Sy1 by a factor of $\sim$ 4.
In addition, a population synthesis modeling of Cosmic X-ray Background (Gilli et al., 1999) indicates that 80\% of the AGN need to be
obscured to produce the CXB spectrum, which is harder than the spectrum of unobscured AGN.

The local background has been subtracted from the PDS data and thus the effect of random AGN and QSO in a given line of sight should have
been removed from our results.
However, there is evidence that the AGN density {\it inside} clusters is enhanced by a factor of 2
compared to non-cluster fields (Molnar et al. 2002; Cappi et al., 2001; Sun \& Murray 2002), perhaps due to galaxy-galaxy interactions.
Their contribution is not removed by the standard background subtraction and thus we need to estimate the number of the excess AGN in
clusters (compared to blank fields). The unobscured AGN are optically identifiable and soft X-ray bright. Thus, it is feasible to find
them from optical catalogs and soft X-ray images, and estimate their contribution to HXR, as we describe in Section \ref{unobscured}.
However, the obscured AGN are a difficult problem for the HXR studies, because the high obscuration by the torus may hide them in the
$<$ 10 keV band, and make the optical detection difficult. At 20 -- 80 keV energies, NH has no effect and the obscured AGN may give a \
significant contribution in this band. We estimate this contribution in Section \ref{obscured}.

\subsection{Unobscured AGN}
\label{unobscured}
A combined ASCA 2 -- 10 keV spectrum of 13 unobscured RIXOS AGNs (Page, 1998) has a photon index of 1.8$\pm$0.1 . Perola et al. (2002)
studied nine bright Sy1 galaxies in the 0.1 -- 200 keV band using BeppoSAX LECS + MECS + PDS data. Their results indicate that in the
20 -- 80 keV band the photon index is 1.8 on average with a standard deviation of 0.1, while all best fit values fall within the range
1.8$\pm0.2$. These two works indicate that a slope of 1.8 is a good representation of unobscured AGN spectra, and that the extrapolation
of the 2 -- 10 keV spectrum up to 80 keV is robust. In our work, when the spectral information is not available, we use a power-law model
with a photon index of 1.8$\pm$0.2 (at a 90\% confidence level) as a reference model to estimate the Sy1 contribution in the PDS 20 -- 80
keV band data.

The variable flux level of Sy1 must be taken into account. A study of 113 Sy1 observed in the ROSAT All Sky Survey and in pointed PSPC and
HRI observations (Grupe et al. 2001) shows that while a few percent of the objects in the sample are transients whose soft band flux
varies by a factor of 100 in timescales of years, $\sim$ 90\% of the AGN vary by less than a factor of 2-3. The hardness ratio analysis is
consistent with no spectral variation. A study of nine Seyfert 1 light curves in the 2 -- 10 keV energy band with RXTE (Markowitz et al.
2001) yields results consistent with variability by less than a factor of 2. Furthermore, they exhibit stronger variability in the
2 -- 4 keV band than in the 7 -- 10 keV band, consistent with the ROSAT study. If this trend continues towards higher energies, variability
by more than a factor of 2 should not be common. Thus in our analysis, when simultaneous normalization level information is not available,
we include $\pm$ 50\% uncertainty (a factor of 3 variation between lower and upper limit at 90\% confidence level) for the AGN contribution
to PDS data.

We searched the SIMBAD database for non-Sy2 AGN within 1.3$^{\circ}$ of the FOV center (we perform a separate treatment for Sy2 in
Section \ref{obscured}). We limited the search to AGN whose redshifts indicate that they belong to the cluster under study.
For each cluster, we studied the 3 best known objects (see Table \ref{agn_tab}).
We also cross-examined the MECS and PSPC images of the clusters in our sample for additional  bright point sources.
Due to a smaller PSF, we used PSPC instead of MECS to identify point sources, and then examined the corresponding MECS image for
excess emission in that sky position. We also examined MECS images for additional variable hard band sources, which were not visible in
PSPC. We assume in the following conservatively that the point sources identified here constitute the excess AGN population inside clusters,
compared to the blank field population, which has not been subtracted from the PDS signal.

The estimation of the AGN contribution to PDS data is difficult since there is no spatially resolved hard X-ray spectroscopic information
for our cluster sample. Thus, where possible, we use the MECS data to obtain the 2 -- 10 keV AGN spectrum and extrapolate it to PDS
energies. We subtract the local background obtained next to the AGN, to ensure similar cluster contributions in both source and background
data. We include the vignetting effect by using ancillary files appropriate for a given off-axis angle as provided by the BeppoSAX team.
This method has the virtue of reducing the uncertainties of the time variability of the AGNs. However, due to the wide PSF of MECS, this
approach is not accurate for the faintest off-axis sources. For those, as well as for the sources outside MECS FOV, we use PSPC 0.4 - 2.0
keV count rates to normalize the reference model, considering the spectral and flux variability as described above.
The details of the AGN modeling in individual cluster fields are given in the Appendix.

\subsection{Obscured AGN}
\label{obscured}
For nearby (z$<$0.1) Sy2 galaxies to significantly affect our results, they need to produce a luminosity of $\sim$ L$_{20-80}$ =
10$^{43-44}$ erg s$^{-1}$ in each PDS pointing (see Section \ref{detect}). A Chandra study of the A2104 cluster
(Martini et al. 2002) revealed five optically unidentified point sources (Sy2 galaxies) whose total luminosity reaches
10$^{43}$ erg s$^{-1}$ in the 20 -- 80 keV band when using a power-law with photon index of 2 to extrapolate from the 2 -- 10 keV band.
This indicates that Sy2s can in principle affect our results.

In the unified scheme of AGN and in the X-ray background synthesis models it is assumed that the intrinsic luminosity distribution of
the obscured and unobscured objects is the same. Assuming further that the relative Sy1-to-Sy2 number densities are similar in the
field and in the cluster environments, we can estimate the 20 -- 80 keV emission of the obscured AGN inside a 
given cluster by multiplying the corresponding Sy1 contribution estimated above by a factor of 4.
The number of galaxies in typical rich clusters is of the order of 100 and thus the assumed number of AGN is only a few per cluster,
which introduces problems of small number statistics. In some clusters there are no catalogued Sy1 and thus no predicted Sy2 signal.
On the other hand, in clusters A1367, A1795, A2029, A2142, A3627, A3667
and Cygnus A the Sy1 based estimate for the non-thermal contribution is higher than the observed non-thermal signal.
Thus, we cannot form a robust Sy2 contamination estimate for each cluster, but rather have to resort to
a sample average Sy1-based estimate for the Sy2 contribution. The estimate is clearly dominated by Cygnus A and
Perseus; excluding these sources, the average Sy2 20 -- 80 keV band luminosity of 4 $\times 10^{43}$ erg s$^{-1}$ is similar to the
average non-thermal luminosity (6 $\times 10^{43}$ erg s$^{-1}$) observed in the sample (see Fig. \ref{f4.fig}). Thus, if the assumptions
involved in the estimation are correct, Sy2 galaxies may potentially be a significant source of contamination in the 20-80 keV band.

The assumption of the similarity of field and cluster point source populations can be addressed studying the Chandra analysis of point
sources in clusters. These observations reach a flux level of 10$^{-15}$ erg s$^{-1}$ cm$^{-2}$ in the 2 -- 10 keV band
(Martini et al. 2002; Sun \& Murray 2002; Molnar et al. 2002). Following the observation  that a substantial fraction ($\sim$ 50\%) of Sy2
are Compton thick (Risaliti et al.), we estimate that the sample average non-thermal luminosity yields absorbed 2 -- 10 kev fluxes of
10$^{-17}$ -  10$^{-13}$ erg s$^{-1}$ cm$^{-2}$, when assuming a power-law with $\alpha_{ph}$ = 2.0 with NH = 10$^{24-25}$ cm$^{-2}$.
Thus, Chandra is sensitive enough to probe a significant fraction of the predicted obscured AGN in clusters. However, 
in several clusters observed with Chandra (e.g. Molnar et al. 2002), no such sources were found.
Also, the above works indicate that the faint point sources in different clusters are of
different nature, and thus do not support the above assumption of substantial field-like Sy2 population in all clusters. Consequently,
the sample average Sy2 contamination level can only be taken as qualitative. In this work we use the quantitative predictions for the flux
of unobscured AGNs, and discuss the possible effects of obscured AGNs in the conclusions.

\section{Results}
\subsection{Detections}
\label{detect}
\begin{figure}
\plotone{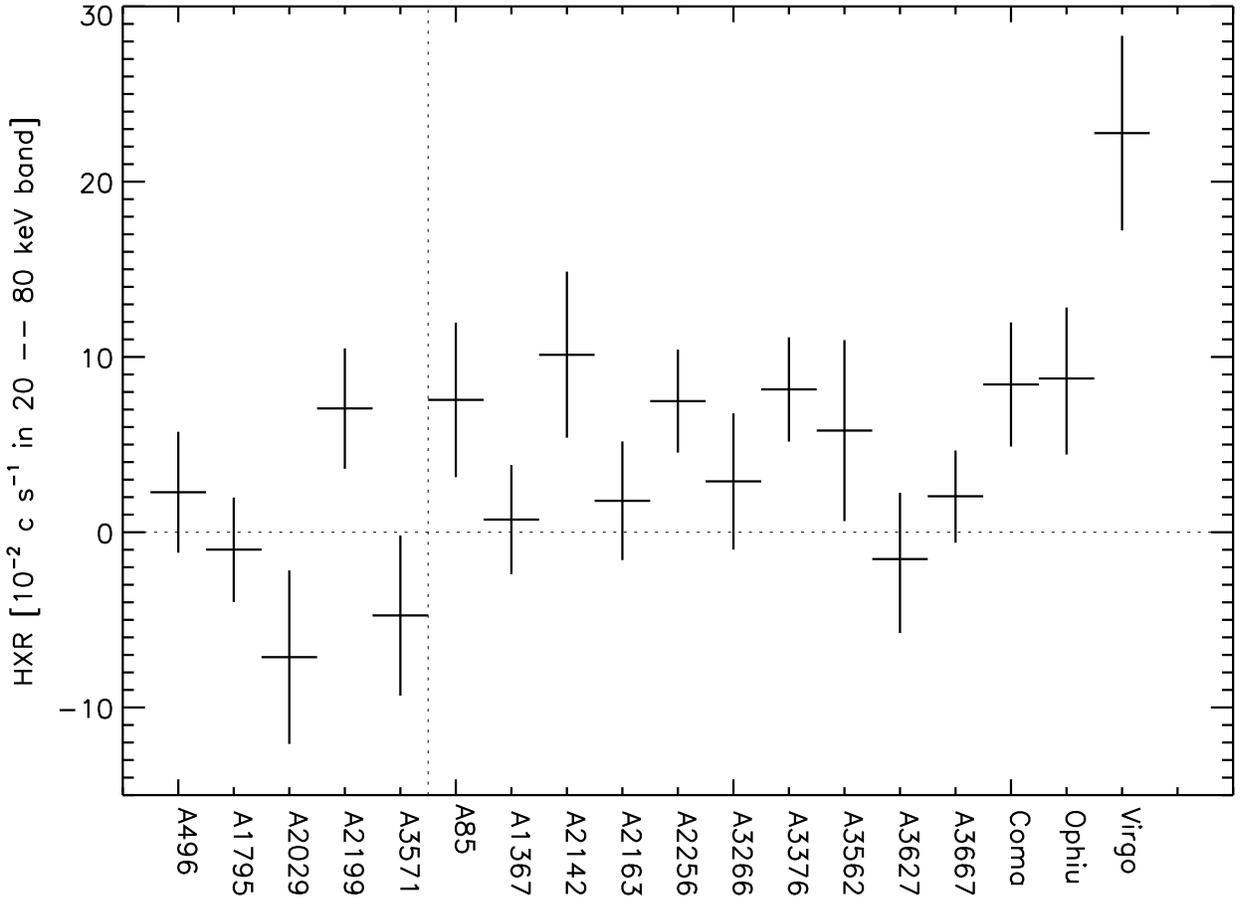}
\caption[]{The non-thermal signal and 1$\sigma$ uncertainties in PDS 20 -- 80 keV band after subtraction of the contributions from the
background, thermal gas and AGN in the field, and after propagating uncertainties due to these subtractions. The dotted vertical line
separates the relaxed clusters (left) from the rest (right).}
\label{f3.fig}
\end{figure}

\begin{figure}
\plotone{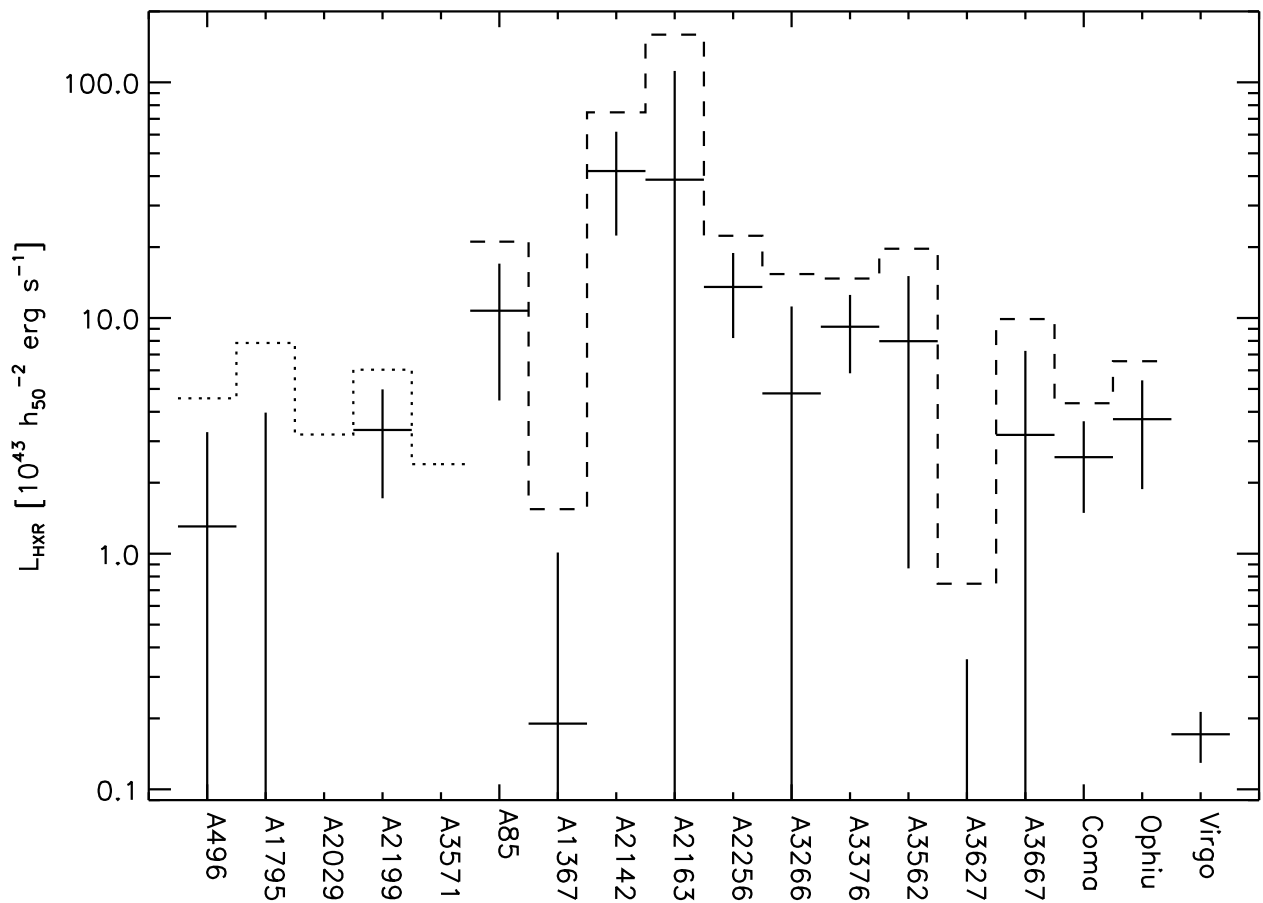}
\caption[]{The luminositites of the non-thermal emission in PDS 20 -- 80 keV band at 1$\sigma$ confidence level, obtained by using a
power-law model with a photon index of 2.0. The dotted and dashed lines show the allowed 90\% upper limit for HXR luminosity in the
relaxed and merger clusters. Note that A3571 and A2029 values are negative at upper 1$\sigma$ level and thus excluded from the
logarithmic plot.}
\label{f4.fig}
\end{figure}

In order to obtain the count rates of the non-thermal emission in 20 -- 80 keV band (HXR) we subtracted the estimated thermal emission and
the unobscured AGN contribution from the PDS data, and propagated the uncertainties arising from background fluctuations, PDS data
statistics, and modeling of the thermal and AGN contributions (see Fig. \ref{f3.fig} and Table \ref{results.tab}). HXR fluxes vary between
0 and 0.1 c/s in 20 -- 80 keV PDS band.
The statistical uncertainties are similar for different clusters, because the PDS signal is dominated by the background and the exposure
times are similar within the sample. The systematic uncertainties due to background fluctuations are comparable to the statistical ones
and common for all clusters. Thus the uncertainties of the background-subtracted PDS count rates are similar in different clusters.
The relative total uncertainties are quite large, ranging from 10\%  to several 100\%. The largest errors
correspond to those clusters with significant Sy1 contaminations.

In the sample there are 15 clusters whose 20 -- 80 keV band signal is
not significantly contaminated by Sy1 (e.g., less than 15\% of the total signal, thus smaller than the statistical errors):
A85, A496, A1795, A2029, A2142, A2163, A2199, A2256, A3266, A3376, A3562, A3571, Coma, Ophiuchus and Virgo.
In $\sim$50\% of these, the non-thermal component is detected at 2 $\sigma$ level (A2142, A2199, A2256, A3376, Coma, Ophiuchus and Virgo).
The 4 $\sigma$ detection of the Virgo cluster constitutes a separate case. Virgo is the nearest cluster, which renders its data of high
S/N, and it features the coolest ICM ($\sim$2 keV), thus giving the least thermal contribution in the PDS band.
Furthermore, the HXR luminosity of Virgo is one order of magnitude smaller than the other 2 $\sigma$ detected clusters, and therefore its
hard excess is more easily produced by unseen AGNs. We confirm the previously published HXR detections of Coma (Fusco-Femiano et al. 1999)
and A2256 (Fusco-Femiano et al. 2000), albeit at lower confidence level due to the level of systematic uncertainties of our work. The HXR
detection of A2199 would be higher than 2.1 $\sigma$, if we assumed that the steep PSPC spectrum of the AGN in the field of A2199 was to
be extrapolated to PDS energies, as in Kaastra et al. (1999). Again due to our AGN modeling, our detection of A3667 is of lesser
significance than that of Fusco-Femiano et al. (2001).

All the clusters detected at 2$\sigma$ level, except A2199, exhibit some degree of merger signatures, i.e. deviations from the
azimuthally symmetric brightness and temperature distributions, reported as follows: A2142 and A3376 (Markevitch et al. 1998),
A2256 (Molendi et al. 2000), Coma (Arnaud et al. 2001), Ophiuchus (Watanabe et al. 2001) and  Virgo (Shibata et al. 2001).
The well established relaxed clusters A1795, A3571 (Markevitch et al. 1998),
A496 (Markevitch et al., 1999b) and A2029 (Sarazin et al. 1998) exhibit less significant detections.
Thus, we divide our sample into two groups: relaxed clusters (A496, A1795, A2029, A2199, A3571) and merger clusters
(A85, A1367, A2142, A2163, A2256, A3266, A3376, A3562, A3627, A3667, Coma, Ophiuchus), excluding Virgo, CygnusA and Perseus, as
explained above.
Assuming that the clusters in both groups lie at the group average redshift and that the intrinsic emission
models are identical inside a group, we formed a weighted mean of HXR and its uncertainty for both groups.
This yields 0.5 $\times$ 10$^{-2}$ c s$^{-1}$ and 4.8 $\times$ 10$^{-2}$ c s$^{-1}$ in PDS 20 -- 80 keV band for the relaxed and
merger group, respectively, i.e. the count rate of the merger group is 10 times as high as that of the relaxed group.
The average redshift of
the relaxed group is lower (0.048) than that of the merger group (0.058), indicating that the higher count rate of the merger group is
not due to a distance effect, but rather that there are intrinsic differences between the two groups.
Due to the large systematic uncertainties due to background fluctuations (1.9--2.7 10$^{-2}$ c s$^{-1}$)
the detection significance of the merger group remains at 2.5$\sigma$, while the relaxed
group count rate is consistent with zero.

To address the intrinsic emission we assumed that it may be modeled with a power-law model with a photon index of 2.0, and
that only the normalization varies between clusters. We normalized this model to the HXR values for each cluster to obtain its luminosity
L$_{HXR}$ in 20 -- 80 keV band at the cluster's  redshift. Luminosities vary from
0 to  10$^{44}$ h$_{50}^{-2}$ erg s$^{-1}$, most values being in the range 10$^{43-44}$ h$_{50}^{-2}$ erg s$^{-1}$ (Fig.\ref{f4.fig} and
Table \ref{results.tab}). Using the above model and the average redshift, we converted the average count rate of the
merger group into luminosity, obtaining 8 $\times 10^{43}$ h$_{50}^{-2}$ erg s$^{-1}$, respectively. Thus, the data are
consistent with a general scenario whereby the relaxed clusters have no HXR component, while merger clusters do, with a
20 -- 80 keV luminosity of $\sim 10^{43-44}$ h$_{50}^{-2}$ erg s$^{-1}$.

\section{Combined spectrum}
\label{comb}
Individual cluster signals are of insufficient S/N for the purpose of constraining the spectral models. Thus, in order to obtain
information of the average non-thermal cluster spectrum, we formed an average cluster spectrum by co-adding bin-by-bin the PDS counts of
each cluster whose 20 -- 80 keV band signal is contaminated by less than 10\% by Sy1 (see \ref{detect}), i.e. A85, A496, A1795, A2029,
A2142, A2163, A2199, A2256, A3266, A3376, A3562, A3571, Coma and Ophiuchus. The co-added exposure time is 560 ks.
To avoid artificial overestimation of the uncertainties, we did not propagate the uncertainties
of the individual spectra, but rather used the combined spectrum to determine the Poissonian uncertainties. The level of systematic
uncertainty due to background fluctuations in the 20--80 keV band is $\sim$15\% of the background-subtracted PDS signal. Since we have no
information on the energy dependence of this quantity, we assumed it to
be a constant 15\% in the 12 -- 115 keV band. Combining this with the uncertainty of AGN contamination, we arrive at 20\% systematic
uncertainty, which we use in the following analysis.

\begin{figure}
\plotone{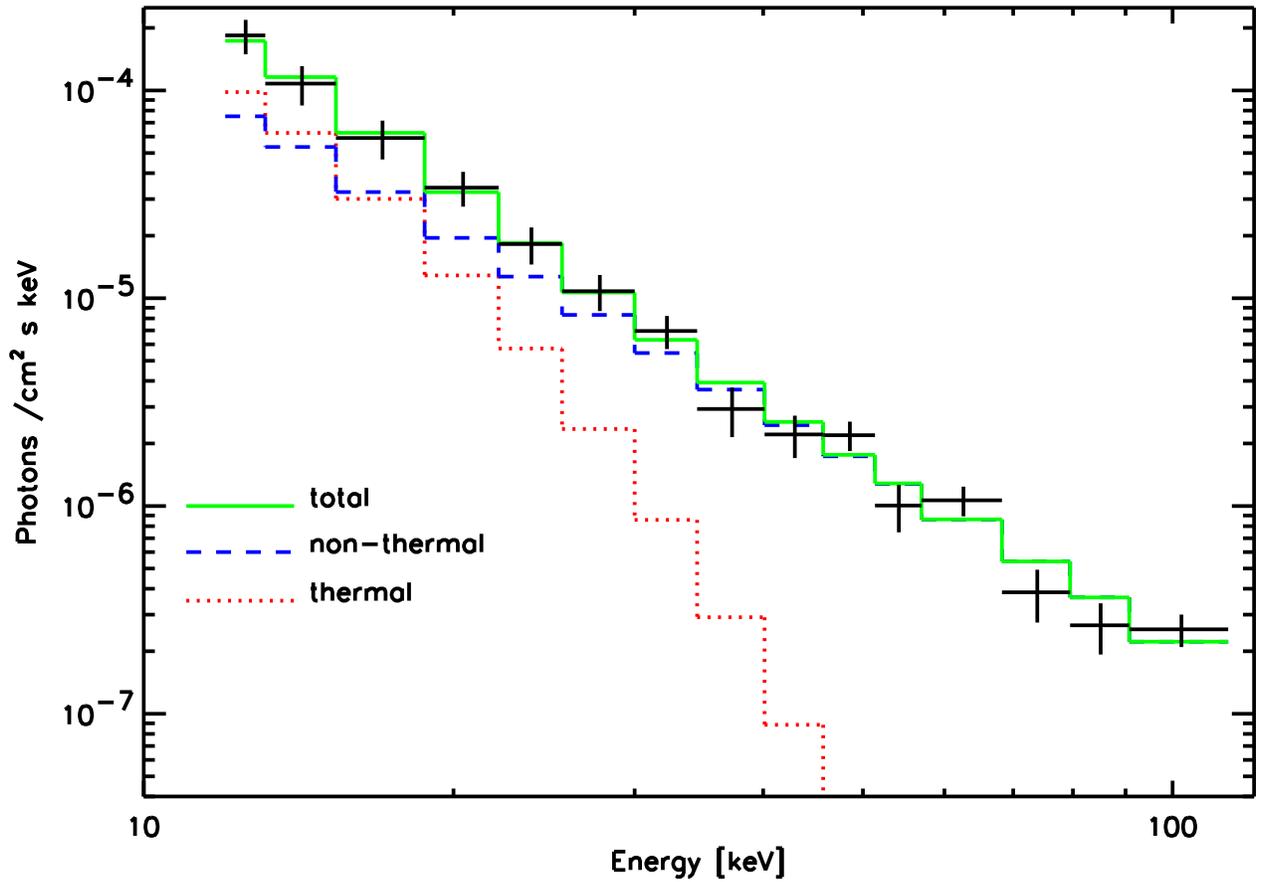}
\caption[]{The combined spectrum of all the clusters not significantly affected by AGN.
The lines show the unfolded model components while the crosses show the data and 1 $\sigma$ errors (including 20\% systematics).
The solid line shows the total model. The dotted line shows the thermal contribution.
The dashed line shows the best fit power-law of $\alpha_{ph} = 2.8$.}
\label{f5.fig}
\end{figure}

In an attempt to account for the total thermal contribution, we first fitted the 12 -- 20 keV band data with a mekal model keeping metal
abundance at 0.3 Solar and the redshift at the median of 0.06. In this band, the typical NH of $10^{20}$ cm$^{-2}$ has no effect, and
thus we exclude the absorption from the model. The best fit temperature is consistent with the median  of 7.8 keV of the sample, implying
that the non-thermal emission does not dominate in the 12 -- 20 keV band. The
non-thermal excess on top of the thermal model is clearly evident: at 100 keV, the thermal model underpredicts the signal by 4 orders of
magnitude (see Fig \ref{f5.fig} for the thermal contribution in the final best fit model).
Fitting the 12 -- 115 keV band data with only a mekal model, we obtain a statistically unacceptable fit with unrealistically
high temperature of 26 keV. This further confirms the existence of an additional hard X-ray component.

We introduced a power-law component to the 12 -- 115 keV band fit, allowing the photon index and the normalization to vary, together
with the mekal temperature and normalization. The best-fit is formally acceptable, with $\chi^{2}$/dof of 10.9/11, yielding a photon index
of 2.8$^{+0.3}_{-0.4}$ (Fig. 7). The typical AGN photon index of 1.8 is ruled out at 98\% confidence level, which argues
against significant Sy2 contamination in the hard X-ray signal.

However, the non-thermal emission in this model in the 12 -- 20 keV band is high, $\sim$ 50\% of the
total. In order to study the relative contribution of the thermal and non-thermal components, we examined the central 8$'$ MECS data of
the largest contributors to the thermal emission in the sample, i.e. Coma and Ophiuchus by fitting the MECS 2 -- 10 keV data with a mekal
+ a power-law with $\alpha_{ph}$ fixed to 2.8. The allowed 1$\sigma$ upper level for the non-thermal contribution, extrapolated to
12 -- 20 keV band, is below 1\%. Thus, the above best fit model requires that the non-thermal emission is extended and negligible in the
central 8$'$.
To confirm that this is the case, one needs to perform spatially resolved hard X-ray spectroscopic analysis on the cluster sample,
which is currently not possible.

Alternately, if we assume that most of the HXR originates from cluster centers, the MECS data require a harder spectrum for it:
Fixing $\alpha_{ph}$ to smaller values, and keeping NH at Galactic values, the mekal + power-law fit to 2--10 keV MECS data of Coma and
Ophiuchus allow bigger contribution from the non-thermal model in the 12 -- 20 keV band. Also, forcing the non-thermal component in the
PDS 12 -- 115 keV fit
to be harder, decreases the non-thermal contribution in 12--20 keV PDS band and with
$\alpha_{ph} \le 1.5$ the non-thermal flux at 12 -- 20 keV  in the best-fit PDS model is below 10\% of the thermal,
consistent with the MECS data of Coma and Ophiuchus. On the other hand, the decreasing non-thermal contribution in the 12 -- 20 keV band
requires higher
temperatures for the best-fit PDS model, and at $\alpha_{ph} \le 1.3$ it exceeds the highest temperature of the cluster sample.
Thus, assuming that the non-thermal emission comes mainly from the cluster centers, its photon index is limited within 1.3--1.5.
However, such hard slopes yield poor fits to the PDS data. Keeping $\alpha_{ph} \equiv 1.3$ the model has
$\chi^{2}$/dof = 20.9/12 and systematically exceeds the data by 20-40\% above 70 keV energies.
Note that assuming an obscured AGN model with $\alpha_{ph}$ = 1.8 and NH = $10^{25}$ cm$^{-2}$ would
yield consistence between MECS and PDS, since the high absorption in 2-10 keV band would hide the non-thermal contribution from MECS data,
but would allow it to dominate in PDS band. However, this forced fit is also significantly worse ($\chi^{2}$/dof = 17.2/12) than that of the
free $\alpha_{ph}$ fit.
Thus the data indicate that the steeper slope (2.4 -- 3.1) and thus the extended distribution of the non-thermal emission are more
likely.
The Sy2 galaxies inside clusters are concentrated in the central high galaxy density regions and thus the indication for extended
nature of the non-thermal emission also argues against the Sy2 origin of the PDS signal.

\section{Models and discussion}
Most models for the HXR emission require acceleration of cluster electrons to
supra-thermal and/or relativistic velocities. Large-scale acceleration is
naturally provided by merger shocks, and our findings (higher HXR detection
significance of the merger clusters compared to relaxed clusters, see Section 5)
are consistent with this basic assumption. A strong merger accelerates electrons
to relativistic velocities and consequently the Cosmic Microwave Background
photons may experience Inverse Compton scattering (IC/CMB) from these
electrons, thus producing hard X-ray emission. Within the framework of the
merger acceleration, the observed photon index (2.4 -- 3.1) of the combined PDS
spectrum in this work implies a power-law form for the differential momentum
spectra of the relativistic ($\sim$ GeV) electrons with a slope of $\mu$ =  3.8
- 5.2. Right after the first acceleration event the primary electron
distribution is predicted to be harder ($\sim$ 2--2.5, see Miniati et al., 2001) 
but the electrons loose energy rapidly and their spectrum in the GeV
range steepens into consistence with that derived from the observed PDS
spectrum.

The IC/CMB model requires a confinement of the relativistic electrons in
clusters, which can be achieved by the cluster magnetic fields. In the presence
of magnetic fields, the relativistic electrons produce synchrotron emission 
at radio wavelengths. Thus, the model naturally predicts a connection
between the non-thermal hard X-rays and radio emission. We have indicated above
a connection between the non-thermal hard X-ray emission and cluster mergers,
which in turn predicts a connection between cluster mergers and radio emission.
Indeed, clusters with a large scale ($>$ 1 Mpc) radio halo possess merger
signatures such as substructure in the X-ray brightness and temperature
distribution and absence of cooling flows (see Feretti 2003 and references
therein). Also, the diffuse radio emission is more common in clusters with
higher X-ray luminosities (Giovannini et al. 1999), perhaps due to energy input
by recent mergers, as in hydrodynamic simulations (Sarazin et al. 2002). Thus,
the observed connections of cluster mergers with radio emission and with
non-thermal hard X-rays support the IC/CMB scenario whereby mergers provide
higher temperatures and luminosities as well as stronger shock acceleration, and
thus stronger radio and non-thermal X-ray emission. Within this framework, the
observed spectral index of the combined PDS spectrum in this work (1.4 -- 2.1)
equals that of the radio spectral index of the synchrotron spectra. Indeed, most
radio-halo cluster observations typically yield spectra with indexes in this
range (e.g. Feretti et al. 2001; Fusco-Femiano et al. 1999), further
strengthening the case for IC/CMB. In a forthcoming paper we will examine the
connection between the HXR and the radio emission in a sample of clusters in
more detail.

The case for IC/CMB is challenged by the clusters A2163, A3266 and A3562
with merger signatures featuring less significant HXR detections from the rest
of the merger group. Also, HXR for a relaxed cluster A2199 is detected with
2$\sigma$ confidence. This implies that the merger is not the only factor
responsible for the non-thermal emission in all clusters, which gives room for
other models. Even though there is a possibility of Sy2 contribution in the PDS
signal (see above), the co-added spectrum is steeper than those observed in AGN
and the indicated extended distribution of HXR is also contrary to the central
concentration of AGN in clusters. Thus the current data argue against
significant contamination by obscured AGN in our sample. The non-thermal
bremsstrahlung model (e.g., Sarazin \& Kempner 2000, Dogiel 2000) predicts
spectral slopes of HXR consistent with our observations, and thus cannot be
ruled out by the fit to the current data. However, bremsstrahlung is a
very inefficient process (see, e.g., Petrosian 2001, Timokhin et al.
2003), and the huge amount of energy input needed to produce the observed level
of hard X-ray emission is ruled out in cases like Coma by X-ray
observations (Petrosian 2001).

The secondary electron models usually predict a harder ($\alpha_{ph} \sim$
1.5-1.75) IC/CMB spectrum (Colafrancesco \& Blasi 1998; Blasi \&
Colafrancesco 1999; Miniati et al. 2001) than that indicated by the
present observations. If we assume that most of the non-thermal hard X-ray
emission originates from the central regions of the clusters, its spectrum
is required to be hard ($\alpha_{ph} =$ 1.3--1.5, see Section \ref{comb}). 
In this case, the spatially concentrated hard X-ray emission is
consistent with the secondary models, which involve the production of
secondary electrons via collisions of relativistic protons which are bound to
the cluster gravitational potential wells (Colafrancesco \& Blasi
1998). However, this model does not fit well the PDS data at highest energies.

\section{Conclusions}
We have studied the hard X-ray emission in 20 -- 80 keV energy band in a sample
of clusters using the BeppoSAX PDS instrument. After removing the contributions
from the cluster thermal component and from unobscured AGN, in $\sim$50\% of the
mildly AGN-contaminated clusters the non-thermal component is detected at 2
$\sigma$ level, the clusters being  A2142, A2199, A2256, A3376, Coma, Ophiuchus
and Virgo. All the clusters detected at 2$\sigma$ level exhibit some degree of
merger signatures, i.e. deviations from the azimuthally symmetric brightness and
temperature distributions (except for A2199). Averaging the PDS 20 -- 80 keV
count rates of the relaxed and merger clusters obtains a 2.5$\sigma$ detection
for the merger group, while the relaxed group count rate is consistent with
zero. Assuming a power-law emission model with a photon index of 2.0 at
the group average redshifts, the average count rates are consistent with a
scenario whereby the relaxed clusters have no HXR component, while mergers do,
with a 20 -- 80 keV luminosity of $\sim 10^{43-44}$ h$_{50}^{-2}$ erg s$^{-1}$.

The co-added spectrum of our sample yields a best-fit photon index of
2.8$^{+0.3}_{-0.4}$ for the non-thermal emission in 12 -- 115 keV band, and we
find indication that it has extended distribution. These indications argue
against significant contamination from obscured AGN, which have harder spectra
and centrally concentrated distribution.

The indicated connection between cluster mergers and the non-thermal hard X-ray
emission is consistent with the Inverse Compton scattering of the Cosmic
Microwave Background photons with merger--accelerated population of relativistic
electrons. In this framework, the observed photon index 
is consistent with a scenario in which a strong acceleration event and
consequent strong IC/CMB energy losses take place. In this
scenario the measured hard X-ray slope corresponds to a differential momentum
spectra of the relativistic electrons with a slope of $\mu$ = 3.8 -- 5.0. The
consequent synchrotron emission spectrum expected from the same electron
population has a spectral index of 1.4 -- 2.1, consistent with radio halo
observations of many merger clusters.

The observed slope of the HXR spectrum is also consistent with the
predictions of the non-thermal bremsstrahlung model. Even though this fit
cannot be ruled out by the current data, the bremsstrahlung model
seems to face a strong energetics problem which does not make it a viable
physical scenario.

Assuming that most of the non-thermal signal originates in the central regions
of clusters, the HXR spectrum is forced to be harder, with a slope $\sim$
1.3--1.5, which turns out to be consistent with secondary electron models.
However, this model provides a worse fit to PDS data and is thus
disfavored by the statistical fit over the primary electron IC/CMB
model.

In conclusion, spatially resolved hard X-ray spectroscopy is needed to
disentangle between primary and secondary electron models for non-thermal hard X-ray emission in
clusters of galaxies.

\acknowledgements

The \sax\ satellite is a joint Italian-Dutch programme. We thank the staffs of the \sax\ Science Data and Operations Control Centers for
help with these observations. J. Nevalainen acknowledges an ESA Research Fellowship, and a NASA grant NAG5-9945. We thank Drs. T. Clarke,
D. Harris, M. Markevitch, M. Page and H. Tananbaum for useful comments and Dr. A. Parmar for his help on the project. M. Bonamente gratefully acknowledges NASA
for support.

\clearpage

\clearpage

\clearpage

\clearpage

\clearpage

\clearpage

\begin{deluxetable}{llllllllll}
%\tabletypesize{\scriptsize}
\tablecaption{Results for HXR}
\tablewidth{0pt}
\tablehead{
\colhead{name}    & \colhead{PDS CR$^{a}$}         & \colhead{CL$_{det}$} & \colhead{thermal$^{a}$}        &
\colhead{AGN\tablenotemark{a}} & \colhead{HXR\tablenotemark{a}} & \colhead{CL$_{HXR}$} & \colhead{HXR$_{90}$\tablenotemark{a}}
& \colhead{L$_{HXR}$\tablenotemark{b}} & \colhead{L$_{HXR90}$\tablenotemark{b}} \\
        & & $\sigma$   & &            &           & $\sigma$ &                            }
\startdata
A85   & 11.6$\pm$4.3 {\bf (-)}  & 2.7  & 4.1$\pm$0.8          & -                     & 7.6$^{+4.4}_{-4.4}$  & 1.7 & 14.8 & 10.7$^{+6.3}_{-6.3}$ & 21.1  \\
A348  &-1.6$\pm$3.7             & ...  &                      &                       &                      &     &      &                      &       \\
A496  & 3.7$\pm$3.4             & 1.1  & 1.5$\pm$0.4          & -                     & 2.3$\pm$3.5          & 0.7 & 8.0  & 1.3$\pm$2.0          & 4.6   \\
A1367 & 2.1$\pm$3.1             & 0.7  & 0.4$\pm$0.2          & 1.0$^{+0.4}_{-0.3}$   & 0.7$\pm$3.1          & 0.2 & 5.9  & 0.2$\pm$0.8          & 1.6   \\
A1750 & 1.6$\pm$4.0  {\bf (-)}  & 0.4  &                      &                       &                      &     &      &                      &       \\
A1795 & 3.0$\pm$2.9             & 1.0  & 3.7$^{+0.7}_{-0.6}$  & 0.3$\pm$0.1           & -1.0$\pm$3.0         & ... & 3.9  & -2.0$\pm$6.0         & 7.8   \\
A2029 & 2.0$\pm$4.8             & 0.4  & 8.9$\pm$1.3          & 0.2$\pm$0.1           & -7.1$\pm$5.0         & ... & 1.0  & -22.0$\pm$15.3       & 3.2   \\
A2142 & 25.0$\pm$4.4 {\bf (+)}  & 5.7  & 11.9$^{+1.6}_{-1.8}$ & 3.0$^{+0.8}_{-0.6}$   & 10.1$\pm$4.8         & 2.1 & 18.0 & 42.0$\pm$19.7        & 74.6  \\
A2163 & 8.3$\pm$3.3             & 2.6  & 6.5$\pm$0.9          & -                     & 1.8$\pm$3.4          & 0.5 & 7.4  & 38.6$\pm$73.1        & 159.4 \\
A2199 & 9.2$\pm$3.4             & 2.7  & 1.6$\pm$0.4          & 0.6$^{+0.4}_{-0.3}$   & 7.1$\pm$3.4          & 2.1 & 12.7 & 3.4$\pm$1.6          & 6.0   \\
A2256 & 11.6$\pm$2.9            & 4.0  & 4.1$\pm$0.5          & -                     & 7.5$\pm$2.9          & 2.5 & 12.3 & 13.6$\pm$5.3         & 22.3  \\
A2390 & 9.4$\pm$4.4  {\bf (-)}  & 2.1  &                      &                       &                      &     &      &                      &       \\
A3266 & 9.5$\pm$3.8             & 2.5  & 6.4$\pm$1.0          & 0.2$^{+0.1}_{-0.1}$   & 2.9$\pm$3.9          & 0.8 & 9.3  & 4.8$\pm$6.4          & 15.4  \\
A3376 & 9.5$\pm$3.0             & 3.2  & 0.4$\pm$0.2          & 0.9$^{+0.3}_{-0.2}$   & 8.2$\pm$3.0          & 2.7 & 13.1 & 9.2$\pm$3.4          & 14.7  \\
A3562 & 6.8$\pm$5.1  {\bf (-)}  & 1.3  & 0.9$\pm$0.6          & -                     & 5.8$\pm$5.2          & 1.1 & 14.3 & 8.0$\pm$7.1          & 19.7  \\
A3571 & -1.7$\pm$4.6 {\bf (+)}  & ...  & 3.0\tablenotemark{c} & -                     & -4.8$\pm$4.6         & ... & 2.8  & -4.1$\pm$3.9         & 2.4   \\
A3627 & 11.4$\pm$3.4            & 3.3  & 6.8$\pm$0.6          & 6.1$^{+2.4}_{-1.5}$   & -1.5$^{+3.8}_{-4.2}$ & ... & 4.7  & -0.2$^{+0.6}_{-0.7}$ & 0.8   \\
A3667 & 8.2$\pm$2.6             & 3.2  & 4.4$\pm$0.4          & 1.7$^{+0.6}_{-0.4}$   & 2.1$\pm$2.7          & 0.8 & 6.4  & 3.2$\pm$4.1          & 9.9   \\
Coma  & 40.2$\pm$3.4            & 11.8 & 30.9$\pm$0.8         & 0.9$\pm$0.6           & 8.4$\pm$3.5          & 2.4 & 14.3 & 2.6$\pm$1.1          & 4.4   \\
Cyg A & 58.6$\pm$3.4            & 17.1 &                      & 51.1$^{+11.8}_{-9.1}$ &                      &     &      &                      &       \\
Ophiu & 75.0$\pm$3.9            & 19.3 & 66.2$^{+2.0}_{-1.2}$ & -                     & 8.8$^{+4.1}_{-4.3}$  & 2.0 & 15.5 & 3.7$\pm$1.8          & 6.6   \\
Perseus & 54.7$\pm$3.5            & 15.7 & 37.6$\pm$0.6         &                       &                 &     &      &                      &       \\
PKS07 & 3.5$\pm$3.3             & 1.1  &                      &                       &                      &     &      &                      &       \\
RXJ01 & 0.8$\pm$3.0             & 0.3  &                      &                       &                      &     &      &                      &       \\
RXJ13 & 6.8$\pm$4.4  {\bf (-)}  & 1.5  &                      &                       &                      &     &      &                      &       \\
Virgo & 27.5$\pm$5.5            & 5.0  & 0.3$\pm$0.2          & 4.5$\pm$0.7           & 22.8$\pm$5.6         & 4.1 & 31.9 &  0.17$\pm$0.04       & 0.24  \\
z3146 & 1.4$\pm$4.4             & 0.3  &                      &                       &                      &     &      &                      &       \\
\enddata
\tablenotetext{a}{$10^{-2}$ c s$^{-1}$ in PDS 20 -- 80 keV band}
\tablenotetext{b}{10$^{43}$ h$_{50}^{-2}$ erg s$^{-1}$ in 20 -- 80 keV band}
\tablenotetext{c}{predicted}
\tablecomments{The count rates (PDS CR) are obtained with PDS in 20 -- 80
keV band using standard background subtraction method (default) or using only pos (+) or neg (-) background pointing and correcting for the systematic effect. The errors include both statistical and systematic uncertainties at 1 $\sigma$ level. CL$_{det}$ gives the confidence level of source detection. ``thermal'' gives the thermal model prediction in 20 -- 80 keV band, normalized to PDS 12 -- 20 keV data, together with 1 $\sigma$ uncertainty due to PDS photon statistics in the 20 - 80 keV band.  ``AGN'' gives the estimated AGN contribution to PDS 20 -- 80 keV band.
HXR and CL$_{HXR}$ give the non-thermal, AGN subtracted count rate with 1 $\sigma$ errors, and its confidence level.
HXR$_{90}$ gives the 90\% confidence upper limit of the count rate of the non-thermal emission.
L$_{HXR}$ gives the luminosity of the AGN subtracted non-thermal component, obtained by
normalizing a power-law with photon index $\alpha_{ph}$ = 2.0 to HXR, and its 1$\sigma$ uncertainties.
L$_{HXR90}$ gives  the the 90\% confidence upper limit of the HXR luminosity.}
\label{results.tab}
\end{deluxetable}

\clearpage

\begin{deluxetable}{lllllllllll}
%\tabletypesize{\scriptsize}
\tablecaption{Thermal models}
\tablewidth{0pt}
\tablehead{
\colhead{name} & \colhead{z} & \colhead{\NH} & \colhead{I$_{1}$/I$_{2}$} & \colhead{r$_{c1}$} &
\colhead{r$_{c2}$} & \colhead{$\beta$} & \colhead{L$_{78'}$\tablenotemark{a}} & \colhead{T}     & \colhead{ab}  &  \colhead{Refs.}  \\
     &   &     &                 & [$'$]    & [$'$]    &         &   & [keV]  & [solar] & $\beta$,T,L}
\startdata
A85              & 0.052   & 2.7   & 0.08 & 3.9  & 0.72 & 0.66       & 10.3  & 6.9$^{+0.4}_{-0.4}$  &                                           & 1,8 ,20   \\
A348             & 0.274   & 3.1   & \multicolumn{4}{c}{STANDARD}    & 3.7   & 4.3$^{+1.6}_{-0.8}$  &                                           & 7,12,12 \\
A496             & 0.033   & 4.2   & 0.05 & 4.0  & 0.55 & 0.65       & 4.3   & 4.7$^{+0.2}_{-0.2}$  &                                           & 1,10,20 \\
A1367            & 0.021   & 2.4   &      & 10.0 &      & 0.61       & 3.7   & 3.69$^{+0.10}_{-0.10}$ &                                         & 1,9 ,20  \\
A1750            & 0.086   & 2.4   & \multicolumn{4}{c}{STANDARD}    & 3.2   & 4.46$^{+0.24}_{-0.24}$ & 0.30$^{+0.10}_{-0.10}$                  & 7,9 ,13  \\
A1795            & 0.062   & 1.0*  & 0.07 & 3.5  & 0.82 & 0.79       & 12.1  & 7.8$^{+1.0}_{-1.0}$  &                                           & 1,8 ,20 \\
A2029            & 0.077   & 3.2*  & 0.05 & 2.8  & 0.68 & 0.71       & 17.2  & 9.1$^{+1.0}_{-1.0}$  &                                           & 1,8 ,20  \\
A2142            & 0.089   & 4.1*  & 0.07 & 4.8  & 1.2  & 0.79       & 22.5  & 9.7$^{+1.5}_{-1.1}$  &                                           & 1,8 ,20  \\
A2163            & 0.203   & 11.9  &      & 1.6  &      & 0.73       & 43.3  & 11.5                 &                                           & 2,11,20  \\
A2199            & 0.030   & 0.86* & 0.18 & 3.2  & 0.81 & 0.66       & 4.4   & 4.8$^{+0.2}_{-0.2}$  &                                           & 1,10,20 \\
A2256            & 0.058   & 4.5*  &      & 5.3  &      & 0.83       & 8.2   & 6.97$^{+0.12}_{-0.12}$ & 0.26$^{+0.02}_{-0.02}$                  & 1,9 ,20 \\
A2390            & 0.228   & 7.0   &      & 0.47 &      & 0.60       & 26.3  & 9.8$^{+0.8}_{-0.7}$  & 0.3$^{+0.1}_{-0.1}$                       & 3,14,20 \\
A3266            & 0.055   & 1.6*  &      & 5.7  &      & 0.74       & 8.1   & 8.97$^{+0.30}_{-0.30}$  & 0.22$^{+0.03}_{-0.03}$                 & 1,9 ,20 \\
A3376            & 0.046   & 4.4*  &  \multicolumn{4}{c}{STANDARD}   & 3.0   & 3.99$^{+0.13}_{-0.13}$ & 0.23$^{+0.04}_{-0.04}$                  & 7,9 ,20 \\
A3562            & 0.050   & 3.8   &      & 1.2  &      & 0.47       & 7.1   & 5.1$^{+0.3}_{-0.3}$  & 0.39$^{+0.08}_{-0.08}$                    & 1,15,20 \\
A3571            & 0.040   & 4.1*  &      & 2.6  &      & 0.61       & 9.7   & 6.9$^{+0.2}_{-0.2}$  &                                           & 1,8 ,20 \\
A3627            & 0.016   & 21.9  &      & 10.0 &      & 0.56       & 3.8   & 6.28$^{+0.18}_{-0.18}$ & 0.27$^{+0.02}_{-0.02}$                  & 4,9 ,20  \\
A3667            & 0.053   & 4.8   &      & 3.1  &      & 0.54       & 14.4  & 7.0$^{+0.6}_{-0.6}$  &                                           & 1,8 ,20 \\
Coma             & 0.023   & 0.9   &      & 10.1 &      & 0.71       & 8.8   & 8.21$^{+0.16}_{-0.16}$ & 0.21$^{0.03}_{-0.03}$                   & 1,18,20  \\
Cygnus A         & 0.057   & 36.1  &      & 0.17 &      & 0.47       & 15.5  & 6.9$^{+1.5}_{-1.3}$  & 0.67$^{+0.12}_{-0.10}$                    & 1,8 ,8 \\
Ophiuchus        & 0.028   & 20.3* & 0.65 & 5.8  & 1.7  & 0.71       & 13.5  & 9.1$^{+0.6}_{-0.5}$  & 0.49$^{+0.08}_{-0.08}$                    & 1,13,13  \\
Perseus          & 0.018   & 14.8  & 0.02 & 13.1 & 2.0  & 0.75       & 9.6\tablenotemark{b} & 6.33$^{+0.21}_{-0.18}$ & 0.41$^{+0.02}_{-0.02}$   & 1,19,13 \\
PKS0745-191      & 0.103   & 42.4  &  \multicolumn{4}{c}{STANDARD}   &       & 8.5$^{+0.6}_{-0.6}$  & 0.38$^{+0.03}_{-0.03}$                    & 7,17,- \\
RXJ0152.7-135.7  & 0.831   & 1.6   &  \multicolumn{4}{c}{STANDARD}   & 11.6  & 6.5$^{+2.9}_{-2.0}$  & 0.5$^{+0.5}_{-0.4}$                       & 7,16,16\\
RXJ1347.5-1145   & 0.451   & 4.8   &      & 0.14 &      & 0.56       & 58.4  & 14.3$^{+1.8}_{-1.5}$ & 0.5$^{+0.2}_{-0.2}$                       & 5,14,14\\
Virgo            & 0.0036  & 2.5   &      & 2.2  &      & 0.45       & 0.7\tablenotemark{b} & 2.35$^{+0.06}_{-0.06}$ & 0.49$^{+0.06}_{-0.06}$   & 6,13,13\\
z3146            & 0.291   & 3.0   &  \multicolumn{4}{c}{STANDARD}   & 28.3  & 7.3$^{+0.9}_{-0.8}$  & 0.3$^{+0.1}_{-0.1}$                       & 7,14,14\\
\enddata
\tablenotetext{a}{$10^{44}$ h$_{50}^{-2}$ erg s$^{-1}$ in 0.1 - 2.4 keV band}
\tablenotetext{b}{power-law component removed}
\tablecomments{The \NH values [ $10^{20}$ atoms cm$^{-2}$ ] are based on Dickey \& Lockman, except for the ones marked with *
which are taken from fine beam HII survey (thin filter) of Murphy et al. (in prep.)
The $\beta$ model parameter reference (7) corresponds to STANDARD model of  $\beta$ =  2/3 and  r$_{c1}$ = 0.2 h$_{50}^{-1}$ Mpc,
due to lack of proper reference.
The abundances, where marked, are taken from the temperature references, otherwise 0.3 Solar is assumed.
The unabsorbed luminosities L$_{78'}$ [$10^{44}$ h$_{50}^{-2}$ erg s$^{-1}$] in 0.1 - 2.4 keV band are obtained by using the $\beta$ models to
extrapolate the luminosities taken from the references papers.}
\tablerefs{(1) Mohr et al. 1999; (2) Vikhlinin et al. 1999; (3) B\"ohringer et al., 1998; (4) B\"ohringer et al. 1996; (5) Schindler et al., 1997;
(6) B\"ohringer et al. 1994; (7) STANDARD; (8) Markevitch et al. 1998; (9) deGrandi et al. 2001; (10) Markevitch et al. 1999; (11) Markevitch et al. 1996;
(12) Colafrancesco et al. 2001; (13) this work, (14) Ettori et al. 2001; (15) Ettori et al. 2000;  (16) Della Ceca et al. 2000;
(17) de Grandi et al. 1999;  (18) Hughes et al. 1993; (19) Allen et al. 1992; (20) Ebeling et al. 1996}
\label{thermod_tab}
\end{deluxetable}

\clearpage

\begin{deluxetable}{lllllllll}
%\tabletypesize{\scriptsize}
\tablecaption{Observation log and AGN information}
\tablewidth{0pt}
\tablehead{
\colhead{cluster}  & \colhead{exp start}  & \colhead{exp end}    & \colhead{t}
& \colhead{PSPC Seq\_ID} & \colhead{RA}       & \colhead{DEC}       & \colhead{AGN} &
\colhead{type}  \\
         & year-mm-dd & year-mm-dd &  ks          &              & (J2000)  & (J2000)   &                }
\startdata
A85      & 1998-07-18 & 1998-07-20 & 42 & RP800250N00  & 00 41 30 & -09 23 00 & none              &     \\
A496     & 1998-03-05 & 1998-03-07 & 42 & RP800024N00  & 04 33 38 & -13 15 43 & none              &     \\
A1367    & 1999-12-21 & 1999-12-23 & 46 & RP800153N00  & 11 44 29 &  19 50 02 &                   &     \\
         &            &            &    &              & 11 45 05 &  19 36 22 & NGC 3862          & AGN \\
         &            &            &    &              & 11 46 12 &  20 23 28 & NGC 3884          & LINER \\
A1795    & 1996-12-29 & 1996-12-29 & 5  & RP800105N00  & 13 48 50 &  26 35 30 &                   &    \\
         & 1997-08-11 & 1997-08-12 & 13 &              & 13 48 52 &  26 35 34 & PKS 1346+26       & LINER \\
         & 2000-01-26 & 2000-01-28 & 42 &              & 13 48 35 &  26 31 08 & 1E1346+26.7       & Sy1   \\
         &            &            &    &              & 13 43 57 &  27 12 41 & RXJ1343.9+2712    & AGN \\
A2029    & 1998-02-04 & 1998-02-05 & 18 & RP800249N00  & 15 10 56 &  05 44 38 &                   &      \\
         &            &            &    &              & 15 11 41 &  05 18 09 & JVAS B1509+054    & Sy1  \\
         &            &            &    &              & 15 11 34 &  05 45 46 & QSO J1511+057     & AGN  \\
A2142    & 1997-08-26 & 1997-08-28 & 44 & RP800415N00  & 15 58 20 &  27 14 00 &                   &   \\
         &            &            &    &              & 16 02 09 &  26 19 46 & IC 1166           & Sy1  \\
         &            &            &    &              & 15 59 23 &  27 03 37 & QSO B1557+272     & Sy1  \\
         &            &            &    &              & 15 58 29 & 27 17 08  & 1E 1556+27.4      & Sy1   \\
A2163    & 1998-02-06 & 1998-02-07 & 5  & RP800188N00  & 16 15 18 & -06 07 11 & none              &      \\
          & 1998-02-21 & 1998-02-22 & 5  &              & 16 15 18 & -06 07 11 & none              &      \\
          & 1998-02-23 & 1998-02-24 & 15 &              & 16 15 18 & -06 07 11 & none              &      \\
         & 1998-03-03 & 1998-03-04 & 22 &              & 16 15 18 & -06 07 11 & none              &      \\
A2199    & 1997-04-21 & 1997-04-23 & 42 & RP800644N00  & 16 28 38 &  39 33 05 & none              &      \\
A2256    & 1998-02-11 & 1998-02-12 & 24 & RP100110N00  & 17 03 58 &  78 38 31 & none              &     \\
         & 1999-02-25 & 1999-02-26 & 40 &              & 17 03 58 &  78 38 31 & none              &     \\
A3266    & 1998-03-24 & 1998-03-26 & 32 & RP800552N00  & 04 31 21 & -61 26 40 &                   &     \\
         &            &            &    &              & 04 38 29 & -61 47 59 & J043829.3-614759  & Sy1 \\
         &            &            &    &              & 04 33 34 & -60 58 30 & C3266-12          & AGN  \\
         &            &            &    &              & 04 34 40 & -60 54 06 & E3266-3           & AGN   \\
A3376    & 1999-10-17 & 1999-10-19 & 54 & RP800154N00  & 06 01 37 & -39 59 25 &                   &      \\
         &            &            &    &              & 05 58 50 & -40 38 48 & J055850.3-403848  & Sy1  \\
A3562    & 1999-01-31 & 1999-02-01 & 23 & RP800237N00  & 13 33 38 & -31 40 12 &                   &     \\
         &            &            &    &              & 13 37 58 & -31 44 12 & 1E 1335.1-3128    & Sy1 \\
A3571    & 2000-02-04 & 2000-02-06 & 31 & RP800287N00  & 13 47 28 & -32 51 56 & none              &     \\
A3627    & 1997-03-01 & 1997-03-02 & 16 & RP800382A01  & 16 14 22 & -60 52 20 & none              &     \\
A3627    & 1997-02-24 & 1997-02-24 & 13 &              & 16 16 29 & -61 03 16 & none              &     \\
A3627    & 1997-03-06 & 1997-03-06 & 14 &              & 16 15 52 & -60 37 17 & none              &     \\
A3667    & 1998-05-13 & 1998-05-14 & 36 & RP800234N00  & 20 11 30 & -56 40 00 &                   &     \\
         &            &            &    &              & 20 11 59 & -57 05 07 & FRL 339           & Sy1 \\
         & 1999-10-29 & 1999-11-01 & 64 & RP800234N00  & 20 11 30 & -56 40 00 &                   &     \\
         &            &            &    &              & 20 11 59 & -57 05 07 & FRL 339           & Sy1 \\
Coma     & 1997-12-28 & 1997-12-30 & 31 & RP800005N00  & 12 59 35 &  27 56 42 &                   &     \\
         & 1998-01-19 & 1998-01-20 & 11 &              & 13 00 22 &  28 24 03 & X-Comae           & Sy1  \\
         &            &            &    &              & 12 57 11 &  27 24 18 & J125710.6+272418  & Sy1  \\
         &            &            &    &              & 13 01 20 &  28 39 57 & 1E 1258+28.9      & AGN  \\
Cyg A    & 1999-10-27 & 1999-10-28 & 34 & RP800622N00  & 19 59 28 &  40 44 02 &                   &    \\
         &            &            &    &              & 19 59 28 &  40 44 02 & QSO B1957+405     & Sy1 \\
Ophiuchus    & 1999-08-22 & 1999-08-23 & 26 & RP800279N00  & 17 12 26 & -23 22 33 & none              &     \\
Perseus    & 1996-09-19 & 1996-09-21 & 38 & RP800186N00  & 03 19 50 &  41 32 24 &                   &    \\
         &            &            &    &              & 03 19 48 &  41 30 42 & NGC 1275          & Sy1 \\
Virgo    & 1996-07-14 & 1996-07-15 & 12 & RP800187N00  & 12 30 50 &  12 25 19 &                   &     \\
         &            &            &    &              & 12 30 49 &  12 23 28 & M87               & S3 or LIN  \\
\enddata
\tablecomments{For each cluster the times of the start and the end of each exposure are given, together with the
exposure times (t) and PDS pointing coordinates. A3627 has offset pointings and they are listed separately. Also listed are the PSPC
pointings studied in this work. The names and coordinates of 3 most
referenced AGN (excluding Sy2) within 1.3$^{\circ}$ radius
from the PDS pointing centers, found from SIMBAD database, are given.}
\label{agn_tab}
\end{deluxetable}

\clearpage

\appendix

\section{THE DETAILS OF THE MODELING OF THE UNOBSCURED AGN IN INDIVIDUAL CLUSTERS}

{\bf A85, A496, A2163,A2256:} There are no catalogued Sy1 within these clusters, nor strong point sources in MECS or PSPC in the field.

{\bf A1367:} MECS spectrum of AGN NGC 3862 provides constraints on the reference model, yielding a PDS estimate of 1.0$^{+0.4}_{-0.3}$
$10^{-2}$ c s$^{-1}$. This value is consistent with HXR and thus the HXR estimate will be very uncertain. The PSPC count rate of NGC 3884
is negligible ($<$ 1 \%) of that of NGC 3862 and thus will not change the AGN contribution estimate.

{\bf A1795:} There is a Seyfert 1 galaxy 1E1346+26.7 5$'$  off-axis. Normalizing the reference model to MECS data gives a PDS estimate of
0.3$^{+0.1}_{-0.1}$ $10^{-2}$ c s$^{-1}$. In the cluster center there is LINER PKS 1346+26. Its flux estimate cannot be given due to the
projection with the bright cluster center. However, LINERS usually have 2 -- 10 keV luminosities 1-3 orders of magnitude smaller than
classical Seyferts (Terashima et al., 2002). AGN RXJ1343.9+2712 is outside the PSPC image and thus an flux estimate cannot be given.
The field of A1795, together with Coma, is unusual in its large number of AGN/QSO. For A1795 this is probably not a problem though,
because HXR is negative (consistent with 0) and likely not significantly contaminated by any AGN. We will use the 1E1346+26.7 estimate in
the following.

{\bf A2029:} The MECS data of an AGN QSO J1511+057 at 8$'$ off-axis constrain the reference model yielding PDS estimate of
0.2$^{+0.1}_{-0.1}$ $10^{-2}$ c s$^{-1}$. A Sy1 JVAS B1509+054 is located at 29$'$ off-axis and thus outside MECS FOV. It is undetected
in the PSPC, and the upper limit of statistical uncertainties allow 10\% of the PDS estimate of QSO J1511+057 which is negligible.

{\bf A2142:} A Seyfert 1 galaxy IC 1166 is outside the PSPC field and thus the flux estimate cannot be given. 4 $'$ off-axis from the
cluster centre there is a Seyfert 1 galaxy 1E 1556+27.4 whose photon index is $\alpha_{ph}$ = 1.9 as observed with ASCA (Markevitch et
al., 1998). The MECS data gives consistently $\alpha_{ph}$ = 1.8$\pm0.1$. Using this and including the statistical uncertainties of the
MECS data, the extrapolated PDS 20--80 keV count rate is  2.2$^{+0.7}_{-0.5}$ $10^{-2}$ c s$^{-1}$, consistent with PSPC estimate of
2.9 $10^{-2}$ c/s. In A2142 there is another Seyfert 1 galaxy, QSO B1557+272. At 17$'$ off-axis, the source is quite diffuse in MECS, but
the data extracted from a 4$'$ circle around the source still gives adequate constraints on the normalization of the power-law model.
Including statistical uncertainties, the PDS prediction is 0.8$^{+0.3}_{-0.2}$ $10^{-2}$ c s$^{-1}$. Thus the combined AGN contribution
to PDS 20 -- 80 keV band is 3.0$^{+0.8}_{-0.6}$ $10^{-2}$ c s$^{-1}$, or 30\% of the HXR signal.

{\bf A2199:} There are no catalogued Sy1 in A2199, but X-ray imaging reveals a bright Seyfert 1 galaxy, RXS J16290+4007 of redshift 0.3,
located 35$'$ off-axis and thus outside the MECS FOV. Its PSPC spectrum is not consistent with the reference model. The data can be
modeled with a combination of thermal and a power-law model, yielding a temperature of 0.1 keV and a photon index of 2.4$\pm$0.1, which
gives negligible contribution to HXR. Allowing for spectral hardening towards higher energies, we fitted the PSPC data with our reference
model + mekal. The fit is bad, but however yields an estimate for HXR of 0.6$^{+0.4}_{-0.3}$ $10^{-2}$ c s$^{-1}$, 8\% of the HXR,
thereby decreasing the detection confidence slightly. To be consistent with the treatment of Seyfert 1 galaxies in the rest of the
sample, we assume the harder spectrum in the following.

{\bf A3266:} A Sy1 J043829.3-614759 at the edge of PSPC gives constraint to the reference model, yielding a PDS estimate of
0.2$^{+0.1}_{-0.1}$ $10^{-2}$ c s$^{-1}$, 5\% of the HXR. AGN C3266-12 and E3266-3 are outside MECS FOV and undetected in the PSPC.
The upper limit allowed by statistical uncertainties of PSPC data, fitted with the reference model, yields a PDS estimate below 1\% level
of HXR and thus negligible. In the PSPC there is a bright point source 1RXS J043356.7-612909 at 04$^{h}$33$^{m}$56.70$^{s}$,
-61${\degmark}$29$\arcmin$ 09.5$\arcsec$. However it is not visible in MECS, implying that the source is either very soft or variable,
and very faint during BeppoSAX observation. Either way it gives no contribution to PDS. Thus we keep the estimate of J043829.3-614759.

{\bf A3376}
In the field there is a bright point source 1RXJ J060113.0-401643 at 06$^{h}$01$^{m}$32$^{s}$,
-40${\degmark}$16$\arcmin$55.7$\arcsec$, 18$'$ off-axis. There is no information available on its nature. MECS constraints on the spectral
slope are poor, but the data however give good constrain on the normalization of the power-law model, when fixing the photon index to
1.8$\pm$0.2. With this model, we obtain PDS estimate of 0.3$\pm0.1$ $10^{-2}$ c s$^{-1}$. A bright source at off-axis 24$'$ co-incides
with QSO 1WGA J0600.5-3937 and source PKS 0558-396. Using MECS spectrum we obtain PDS estimate of
0.6$^{+0.2}_{-0.2}$ $10^{-2}$ c s$^{-1}$. A Sy1 J055850.3-403848 within A3376 at off-axis of 50$'$ has count rate much below the above
sources and thus has no effect on the combined estimate of 0.9$^{+0.3}_{-0.2}$ $10^{-2}$ c s$^{-1}$, $\sim$10 \% of the HXR.

{\bf A3562:} Sy1 1E1335.1-3128 at 60$'$ off-axis is undetected in PSPC, implying a negligible HXR contribution. A poor cluster
SC1329-313 is included in the PDS FOV.

{\bf A3571:} There are no catalogued AGN or QSO in the cluster. There is a bright point source HD 119756, an X-ray binary HD 119756 in
the field at RA, DEC = 13$^{h}$45$^{m}$41.5$^{s}$, -33${\degmark}$02$\arcmin$32$\arcsec$. It is obscured by the MECS calibration source
and PSPC spectrum indicates thermal spectrum with T = 0.4 keV with no evidence for power-law component. Thus for A3571 we estimate
negligible PDS contribution from point sources.

{\bf A3627:} Close to the edge of MECS there is a projected Seyfert 1 galaxy 1WGA J1611.8-6037. Using the MECS spectrum we normalized the
reference model and obtained PDS estimate of 6.1$^{+2.4}_{-1.5}$ $10^{-2}$ c s$^{-1}$. This is consistent with the HXR estimate which
will thus be very uncertain.

{\bf A3667:} Seyfert 1 galaxy FRL 339 within A3667 is close to the edge of MECS FOV. The MECS spectrum provides constraint for the
power-law component as 1.9$^{+0.2}_{-0.2}$, consistent with the reference model. Using the MECS data we normalized the reference model
and obtained PDS estimate of 1.0$^{+0.4}_{-0.2}$ $10^{-2}$ c s$^{-1}$. In the PSPC image there are two other bright non-catalogued point
sources, 2E 2007.4-5653 at RA, DEC = 20$^{h}$ 11$^{m}$ 28.6$^{s}$, -56${\degmark}$44$\arcmin$13$\arcsec$ and
1RXS J201455.6-565833 at RA, DEC = 20$^{h}$14$^{m}$55.6$^{s}$, -56${\degmark}$58$\arcmin$33$\arcsec$. The former is undetected in MECS
because it is projected at the bright cluster center and the latter is outside MECS FOV. Both are classified as X-ray sources.
The PSPC data of 2E 2007.4-5653 is consistent with the reference model, and gives a PDS estimate of
0.7$^{+0.5}_{-0.3}$ $10^{-2}$ c s$^{-1}$. The PSPC data of 1RXS J201455.6-565833 is not consistent with the reference model, and requires
a steeper photon index. The PDS prediction with this model is insignificant. The combined AGN contribution (using FRL 339 and
2E 2007.4-5653) is 1.7$^{+0.6}_{-0.4}$ $10^{-2}$ c s$^{-1}$, $\sim$ 80\% of HXR. The point source contamination is not discussed in the
report on the marginal hard excess of A3667 (Fusco-Femiano et al. 2001).

{\bf Coma:} The well known Seyfert 1 galaxy X-Comae is just at the edge of the MECS FOV. Fusco-Femiano et al. (1999) used MECS data to
show that the allowed upper flux level for X-Comae is $\sim$ 15 \% of the hard X-ray excess component at 2 -- 10 keV, using a power-law
component with $\alpha_{ph} = 1.8$. We used PSPC data to check the normalization of the reference model. Including the spectral and flux
level variation uncertainties we obtain a PDS 20 -- 80 keV estimate of 0.9$^{+0.6}_{-0.6}$ $10^{-2}$ c/s or 10 \% of the 20--80 keV HXR
emission, consistent with Fusco-Femiano et al. (1999). AGN 1E 1258+28.9 at off-axis of 50$'$ is obscured by the PSPC mirror support
structure. The useful data still indicates a similar count rate as for X-Comae, indicating a significant contribution to PDS. However,
the nature of the source is not well known and the extrapolation towards higher energies is not justified. Sy1 J125710.6+272418 is
undetected and thus its contribution is negligible compared to that of X-Comae. Coma field contains an unusually high number, 26, of
catalogued AGN/QSO, perhaps because Coma field is better studied than others. However, to make up all the HXR, seven objects like
X-Comae are needed and this is ruled out for Coma based on the PSPC image.

{\bf Cygnus A:} There is a powerful radio galaxy QSO B1957+405 in the center of Cygnus A. We extracted central 2$'$ MECS spectrum and
modeled it as a sum of mekal and self-absorbed power-law, both absorbed by the galactic \NH. The best fit photon index
$\alpha_{ph}$ = 1.9$^{+0.2}_{-0.2}$ is consistent with the ASCA result (Markevitch et al. 1998), and with our reference model. We thus
used the 2$'$ MECS data to normalize the reference power-law model, including a mekal model with temperature, metal abundance and
normalization as free parameters. Extrapolating the resulting power-law model to higher energies we obtained the PDS estimate of
51.1$^{+11.8}_{-9.1}$ $10^{-2}$ c s$^{-1}$. This is consistent with the total observed PDS emission in this band and thus the HXR
estimate will be uselessly uncertain. We thus reject Cygnus A from further analysis.

{\bf Ophiuchus:} There are no catalogued AGN or QSO in Ophiuchus. In PSPC image there is a bright point source RXS J171209.5-231005 at
RA,DEC = 17$^{h}$12$^{m}$09$^{s}$, -23${\degmark}$09$\arcmin$50$\arcsec$, classified as X-ray source. It is undetected in MECS due to
projected bright cluster center in the line of sight. The PSPC spectrum exhibits a 2 component spectrum, consisting of a thermal one with
T $\sim$ 1 keV and a very steep ($\alpha > 3 $) power-law, which gives negligible contribution to PDS HXR.

{\bf Perseus:} Perseus hosts a well known AGN NGC1275 in the center. HEAO I observations revealed a non-thermal component in Perseus data
at 20 -- 50 keV band  (Primini et al. 1981). The excess was modeled with a power-law model whose best fit photon index $\alpha_{ph}$ is
1.9$\pm$0.3 at 90\% confidence. They also report that the source exhibits no significant variations above 25 keV in time scale of 4
years. PDS data are of high enough quality to perform two component fit, if we fix photon index ($\equiv$ 1.9) and mekal abundance
($\equiv$ 0.3). The resulting temperature 6.3$\pm$0.4 keV is identical with the Ginga value (Allen et al. 1992). The power-law component
has 25 -- 40 keV luminosity of 1.8$\pm0.5$ $10^{43}$ erg s$^{-1}$, 4 times smaller than The HEAO I value, implying variability on a time
scale of 20 years.

Due to the high brightness of cluster thermal emission in the center, compared to that of NGC1275, the central 2$'$ MECS data do not
provide decent constraints on the internal \NH\ or the power-law slope of the AGN. However, modeling the central MECS data with
mekal + the above power-law component reveals that the power-law model given by PDS data contributes only a few \% of the total emission.
This component modifies the total model only slightly and the fit is acceptable. If let free, the allowed upper limit for the
normalization of the power-law component is 3 times as high as the best value given by PDS data. Thus, the data are consistent with all
of the non-thermal emission coming from NGC1275. We thus reject Perseus from further analysis.

{\bf Virgo:} Virgo has an active nucleus M87 and a jet in the center. XMM-Newton data yields power-law slopes of
2.2$\pm$0.2 and 2.5$\pm$0.4 for the nucleus and the bright knot in the jet (B\"ohringer et al. 2001) at 90\% confidence, and no
indication of excess absorption. The 2$'$ MECS data do not provide good constraint on the slope of the non-thermal component. We thus fit
the 2$'$ MECS data with a model consisting of mekal and power-law, both absorbed by the galatic \NH, fixing the photon index to 2.3,
based on the XMM-Newton observations (B\"ohringer et al. 2001), thus obtaining the normalization and its uncertainty for the power-law
component. We determined the thermal component of Virgo by using the above determined central power-law model together with a mekal when
fitting 0--8$'$ keV MECS data. The best fit parameters T = 2.35$\pm$0.04 keV and abundance 0.49$\pm$0.04 Solar are consistent with
XMM-Newton results. Letting only the thermal model normalization as a free parameter, we then normalized this model to PDS FOV using
12--20 keV PDS data. According to this model, M87 contributes 17$\pm$3\% of the non-thermal emission in 20 -- 80 keV band.
Allowing spectral variability for M87, we repeated the above exercise keeping $\alpha_{ph}$ at 2.0 and 1.7. The resulting M87
contribution to HXR is 20 -- 30\% and 30 -- 50\%, respectively. Unless the spectrum of M87 has a strong hard excess, the non-thermal PDS
signal of Virgo can not be explained entirely by M87. In the following we keep the $\alpha_{ph}$ = 2.3 results, i.e. M87 contribution of
4.5$\pm$0.7 $10^{-2}$ c s$^{-1}$, and the thermal model prediction of 0.3$\pm$0.2$10^{-2}$ c s$^{-1}$ to the PDS 20 -- 80 keV band.

\end{document}